\documentclass[a4paper,12pt]{article}

\usepackage{amsmath}
\usepackage{amsfonts}
\usepackage{amssymb}
\usepackage{graphicx}
\usepackage{dcolumn}
\usepackage{bm}
\newcommand{\bx}{{\bf x}}

\newcommand{\ve}{{\bf e}}

\newcommand{\dd}{{\rm d}}
\newcommand{\ii}{{\rm i}}

\newcommand{\mR}{{\cal R}}
\newcommand{\mS}{{\cal S}}
\newcommand{\mD}{{\cal D}}

\newcommand{\en}{e}
\newcommand{\ek}{e_r}

\newcommand{\ze}{\zeta}
\newcommand{\zeb}{\overline{\zeta}}
\newcommand{\zb}{\overline{z}}
\newcommand{\ylm}{Y_{l}^m(\theta,\varphi)}
\newcommand{\yslm}{\,_{s}Y_{l}^m(\theta,\varphi)}

\newcommand{\ymslmm}{\,_{-s}Y_{l}^{-m}(\theta,\varphi)}

\newcommand{\yslmvp}{\,_{s}Y_{l}^m(\theta',\varphi')}
\newcommand{\ylmp}{Y_{l'}^{m'}(\theta,\varphi)}

\newcommand{\yslmp}{\,_{s}Y_{l'}^{m'}(\theta,\varphi)}
\newcommand{\ylmc}{{Y_{l}^m}^*(\theta,\varphi)}
\newcommand{\yslmc}{\,_{s}{Y_{l}^m}^*(\theta,\varphi)}
\newcommand\spart{\;\raise1.0pt\hbox{$/$}\hskip-6pt\partial}
\newcommand\spartb{\;\overline{\raise1.0pt\hbox{$/$}\hskip-6pt
\partial}}

\newcommand{\be}{\begin{equation}}
\newcommand{\ee}{\end{equation}}
\newcommand{\bea}{\begin{eqnarray}}
\newcommand{\eea}{\end{eqnarray}}
\newcommand{\al}{\alpha}

\newcommand{\pa}{\partial}

\newcommand{\ints}{\int_0^{\chi_S}d\chi}

\begin{document}
\def\thefootnote{\fnsymbol{footnote}}

\begin{center}
\Large{\textbf{Full-sky lensing shear at second order}}
\\[0.5cm]
\large{Francis Bernardeau, Camille Bonvin and Filippo Vernizzi}
\\[0.5cm]

\small{
\textit{CEA, Institut de Physique Th{\'e}orique, 91191 Gif-sur-Yvette c\'edex, France\\ CNRS, URA-2306, 91191 Gif-sur-Yvette c\'edex, France}}

\end{center}

\vspace{1cm}

\hrule \vspace{0.3cm}
\noindent \small{\textbf{Abstract}} \\[0.3cm]
\noindent
We compute the reduced cosmic shear up to second order in the gravitational potential without relying on 
the small-angle or thin-lens approximation. This is obtained by solving the Sachs equation which describes the deformation
of the infinitesimal cross section of a light bundle in the optical limit, and maps galaxy intrinsic shapes into their angular images.
The calculation is done in the Poisson gauge without a specific matter content, including vector and tensor perturbations generated at second order and taking account of the inhomogeneities of a fixed redshift source plane. Our final result  is expressed in terms of spin-2 operators on the sphere and is valid on the full sky. 
Beside the well-known lens-lens and Born corrections that dominate on small angular scales, we find new nonlinear couplings. These are a purely 
general relativistic intrinsic contribution, a coupling between the gravitational potential at the source with the lens,
couplings between the time delay with the lens and between two photon deflections, as well as
nonlinear couplings due to the second-order vector and tensor components. The inhomogeneity in the redshift of the source 
induces a coupling between the photon redshift with the lens. All these corrections become important 
on large angular scales and should thus be included when computing higher-order 
observables such as the bispectrum, in full or partially full-sky surveys.
\vspace{0.5cm} \hrule
\def\thefootnote{\arabic{footnote}}
\setcounter{footnote}{0}

%\footnotesize
\parskip 0pt
\vspace{1cm} 

\newpage

%%%%%%%%%%%%%%%%%%%%%%%%%%%%%%%%%%%%%%%%%%%%%%%%%%%%%%%%%%%%%%%%%%%
\section{Introduction}

In the early nineties, cosmic shear  
was predicted to be a promising way to study the distribution of matter
in the Universe \cite{1991MNRAS.251..600B,1992ApJ...388..272K,1991ApJ...380....1M} and
since its first detection \cite{2000MNRAS.318..625B,2000astro.ph..3338K,2000A&A...358...30V,2000Natur.405..143W}
it has shown to be a precious mean of investigations of the large-scale structure of the Universe, enabling us to explore dark energy properties or uncover signatures of mode coupling effects 
\cite{2008ARNPS..58...99H,1999ARA&A..37..127M,2008PhR...462...67M}. 

So far cosmic shear surveys have covered only a limited field in the sky.  
For instance, the CFHTLS,\footnote{http://www.cfht.hawaii.edu/Science/CFHTLS} 
which has produced  very promising results over the last years  (see e.g.~\cite{2008A&A...479....9F} 
for a recent account of these observations), is limited to about a 170 squared degree range. 
With the demonstration of the robustness of cosmic shear observations, (nearly) full-sky surveys such as 
Pan-STARRS, DES, LSST, JDEM, or Euclid are under preparation. They will open the way to new types of studies.
Akin to CMB observations, such surveys will be an excellent tool 
to explore the physics of the Universe at scales comparable to the Hubble radius, therefore testing 
genuinely general relativistic effects.\footnote{See \cite{2009arXiv0907.0707Y} for a recent account of these effects 
on galaxy clustering observations.} 
In particular, the study of mode couplings, already well established on Newtonian scales,
can be extended at these very large scales
therefore testing the details of our understanding of the origin and formation of the large-scale structure.

Such an investigation requires that we know the types of mode couplings that are expected to 
be seen at such large scales. Calculations have been undertaken to predict the nonlinear growth of metric and 
density fluctuations after modes reenter the Hubble radius \cite{Bartolo:2005kv,Boubekeur:2008kn,Fitzpatrick:2009ci}. 
In the context of the CMB anisotropies, progress has been recently 
made in understanding the effect of these nonlinearities, 
from concentrating on the large angular scales \cite{Pyne:1995bs,Bartolo:2004ty,2009JCAP...08..029B} 
to the details of the physics of recombination (see for instance \cite{2006JCAP...06..024B,2009CQGra..26f5006P}).

So far the investigations of mode couplings in weak lensing  were limited to small angular scales, corresponding to 
scales much smaller than the angular diameter distance at the source. Accidentally, this distance 
roughly corresponds to the Hubble radius at the source. Thus, on these scales one can consistently neglect 
general relativistic effects that are suppressed by the ratio between the scale probed and the Hubble scale.
On small angular scales the dominant contribution to the cosmic shear comes from fluctuations of the 
gravitational potential transverse to the line of sight. Perturbations along the line of sight 
average out and do not yield appreciable effects.
In this regime the dominant geometrical mode couplings were 
identified more than a decade ago in \cite{1997A&A...322....1B}. 
They include the Born correction and the lens-lens coupling. 
In the so-called Born approximation one integrates the lensing distortion 
over an unperturbed photon path. One can consider the correction due to the fact that the photon path is 
perturbed. The lens-lens coupling consists in the correction due to the deformation of a distant lens 
caused by a foreground one. 
The consequences of these effects have been extensively described  in the literature
and they have been found to have an impact on both the shear power spectrum 
and higher-order statistical observables such as the bispectrum \cite{Cooray:2002mj,2003MNRAS.344..857T,Dodelson:2005zj,2006JCAP...03..007S,Hilbert:2008kb,Krause:2009yr}.  
As the shape distortion probes the reduced shear rather than the shear itself, there is another correction associated to the nonlinear conversion between these two quantities 
\cite{Dodelson:2005rf,Schneider:1997ge,Krause:2009yr}.
Finally, another nonlinear effect is the source-lens clustering,
due to the fact that
the source of the lensed light is itself a perturbed field with specific clustering properties 
correlated with the lens
\cite{1998A&A...338..375B,2002A&A...389..729S}.
For the current surveys restricted to a limited angular field all
these types of couplings are undoubtedly the dominant ones.

In view of full-sky surveys one needs to go beyond the small-angle approximation and probe 
scales of the order of the angular diameter distance to the source. 
In this case fluctuations along the line of sight are not negligible and terms 
other than those described above may become important. 
Furthermore, as we are probing scales comparable to the Hubble size,
one needs to undertake a full general relativistic treatment. This is 
important in order to compute accessible higher-order observables in full or almost full-sky surveys. In 
particular, in such surveys, it becomes necessary if one wants to compute the lensing bispectrum in the squeezed configuration,
when one of the scales probed is taken to be much larger than the other two.

We present here the exhaustive calculation of the weak lensing cosmic shear at second order 
including all general relativistic contributions and
without relying on the small-angle or thin-lens approximations. However, 
we do not include in our study the effect of source-lens clustering 
\cite{1998A&A...338..375B,2002A&A...389..729S}
and other intrinsic effects in the alignment and ellipticity of galaxies.
We work in the so-called generalized Poisson gauge without specifying the matter content of the Universe. 
We assume that there are no primordial vector and tensor perturbations. However, we will take into account vector and tensor components of the metric generated at second order from scalar fluctuations. 
In this gauge we
derive the reduced -- i.e.~observable -- shear by solving the Sachs equation  \cite{1961RSPSA.264..309S}
which describes the distortion of 
the cross section of an infinitesimal bundle of light rays in the geometric-optics limit. 
The advantage of using the Sachs equation instead of  the geodesic equation is that it deals only with physically observable quantities. As the resolution of the Sachs 
equation is extremely tedious and involves a large number of terms we will develop tests that allow us to 
check the validity of its solution. 
In particular, some of the contributions to the second-order shear that we compute -- and that are usually neglected 
in the small-angle approximation -- can be compared to those expected from the lensing shear at linear order 
in a universe with spatial curvature. 

Using the solution of the Sachs equation we will compute the reduced shear by adding the nonlinear corrections coming from the relation between this quantity and the shear itself.
Finally, as in the Poisson gauge hypersurfaces of constant redshift are inhomogeneous, 
we take into 
account the corrections due to the 
inhomogeneity of the redshift of the source \cite{2008PhRvD..78l3530B}. 
The final observable reduced shear field that we obtain is a gauge invariant quantity, although  its separate contributions are not necessarily so. As it is customary for CMB polarization, we express the reduced shear in terms of spin-2 operators on the sky. In particular,
angular gradients on the sky will be written in terms of spin raising and lowering operators, whose eigenfunctions are 
the well-known spin-weighted spherical harmonics.

The plan of the paper is the following. In Sec.~\ref{sec:weak} we give the outline of our calculation. 
In particular, we describe 
the Sachs equation and how the transverse size of a propagating beam can be related to the observable reduced shear
including the effects of the source redshift inhomogeneities.
We will solve the Sachs equation at first order in Sec.~\ref{sec:linear} while the full second-order calculation
will be presented in Sec.~\ref{sec:shear_2nd}. 
We discuss and comment on our results in Sec.~\ref{sec:conclusion} 
in the context of cosmic shear surveys regarding the generation of $B$ modes 
and the expected contributions to the cosmic shear bispectrum.

%%%%%%%%%%%%%%%%%%%%%%%%%%%%%%%%%%%%%%%%%%%%%%%%%%%%%%%%%%%%%%%%%%%%%%%%%%%%%%%%%%%%%%%%%%%%%%%%%%%%%%%%%%%%%%
%%%%%%%%%%%%%%%%%%%%%%%%%%%%%%%%%%%%%%%%%%%%%%%%%%%%%%%%%%%%%%%%%%%%%%%%%%%%%%%%%%%%%%%%%%%%%%%%%%%%%%%%%%%%%%
%%%%%%%%%%%%%%%%%%%%%%%%%%%%%%%%%%%%%%%%%%%%%%%%%%%%%%%%%%%%%%%%%%%%%%%%%%%%%%%%%%%%%%%%%%%%%%%%%%%%%%%%%%%%%%

\section{The weak lensing equations}
\label{sec:weak}

We are interested in studying the propagation of a light bundle, i.e.~a collection of nearby light rays 
\cite{1961RSPSA.264..309S,wald,SEF,1994CQGra..11.2345S,Lewis:2006fu,Uzan:2000xv}.
We consider two nearby null geodesics $x^\mu(\lambda)$ and $x^\mu (\lambda) + \xi^\mu(\lambda)$ that lie in the past-light cone 
of an observer $O$, connected by a deviation vector field $\xi^\mu$. The affine parameter $\lambda$ 
is chosen in such a way that it assumes the same value at $O$ for all geodesics, i.e.~$\lambda_O=0$. Thus, at the observer $\xi^\mu(0) =0$. 
We denote by $k^\alpha={\dd x^\alpha}/{\dd \lambda}$ the wave vector of the photons. This
obeys the geodesic equation,
\be
\label{geodesic}
\frac{D k^\mu}{D \lambda}=0~,
\ee
where  $\frac{D}{D\lambda}\equiv k^\al\nabla_\al$ is the covariant derivative along the geodesic.
For rays with an infinitesimal separation the connecting vector $\xi^\mu$ is also infinitesimal and lies on the null surface, 
i.e.~$\xi^\mu k_\mu=0$ everywhere along the geodesics. The evolution equation of the connecting vector reads $D \xi^\mu /D\lambda = \xi^\nu \nabla_\nu k^\mu$. Indeed, in some arbitrary coordinate system one has
\bea
 k^\nu \nabla_\nu \xi^\mu & =& \frac{\dd}{\dd \lambda} \xi^\mu + \Gamma_{\alpha \beta}^\mu k^\alpha \xi^\beta \nonumber \\ 
 &=& 
 k^\mu (x+ \delta x)  - k^\mu(x) + \Gamma_{\alpha \beta}^\mu k^\alpha \xi^\beta \nonumber \\ &=&
   \xi^\nu \partial_\nu k^\mu + \Gamma_{\alpha \beta}^\mu k^\alpha \xi^\beta  \;.
\eea
By taking the covariant derivative of this equation along the photon geodesic and using the geodesic equation (\ref{geodesic}) one obtains the Sachs equation \cite{1961RSPSA.264..309S},
\begin{equation}
\label{sachs}
\frac{D^2 \xi^{\mu}}{D \lambda^2}=R^{\mu}_{\; \; \nu\alpha\beta}\xi^{\beta}k^{\nu}k^{\alpha}\;,
\end{equation}
where we have used the Ricci identity $(\nabla_\alpha \nabla_\beta - \nabla_\beta \nabla_\alpha) k^\mu = R^{\mu}_{\; \; \nu \alpha\beta} k^\nu$ and $R^{\mu}_{\; \; \nu\alpha\beta}$ is the Riemann tensor. This equation describes the evolution of a light bundle along the geodesic.

Let us consider the case of a light beam emitted by a galaxy at spacetime position $S$ and received by an observer at $O$.
We denote by $v^\mu_O$ the observer $4$-velocity. It is convenient to define 
an orthonormal spacelike basis $n_{a}^{\ \mu}$, with $a=1,2$, orthogonal to $k^\mu$ and to the observer velocity $v_O^\mu$ and such that $g_{\mu\nu}n_{a}^{\ \nu}n_{b}^{\ \mu}=\delta_{ab}$.\footnote{More generally, it is possible to introduce an induced 2D metric on the subspace described by $n_{a}^{\ \mu}$ by imposing $g_{\mu\nu}n_{a}^{\ \nu}n_{b}^{\ \mu}=~\!^{\rm 2D}\!\!g_{ab}$. This is particularly convenient when 
employing spherical polar coordinates to describe the 3D space. 
In this case the Latin indices $a,b,\dots$ %could be covariant or contravariant and 
are raised and lowered by the metric $^{\rm 2D}\!g_{ab}$. As here we take $^{\rm 2D}\!g_{ab} = \delta_{ab}$, upper or lower positions of the indices are irrelevant and repeated indices represent a summation over $a=1,2$.}
At the observer position these vectors form a basis,
$\{n_1^{\ \mu},n_2^{\ \mu},k^\mu,v_O^\mu\} $,
which can be parallel transported along the geodesic,
\be
\label{parallel_transport}
\frac{D n_a^{ \ \mu}}{D \lambda}=0, \hspace{0.5cm}\; \qquad
\frac{D v_O^\mu}{D \lambda}=0~.
\ee
The subspace defined by $\{ n_1^{ \ \mu}(\lambda),~n_2^{ \ \mu} (\lambda)\}$ is called the {\it screen} adapted to $v_O^\mu$ and $k^\mu$.

We can write the deviation vector $\xi^\mu(\lambda)$ in this basis. Using the fact that $\xi^\mu k_\mu=0$ along the geodesic, the component along $v_O^\mu$ vanishes and we have, for all $\lambda$,
\be
\label{xibase}
\xi^\mu=\xi^a n_a^{ \ \mu}+\xi^0k^\mu~,
\ee
with $\xi^a(0) = 0$ and $\xi^0(0) =0$ at the observer.
We can then plug this decomposition into the Sachs equation (\ref{sachs}). Using the symmetry properties of the Riemann tensor the right-hand side of this equation becomes $R^{\mu}_{\ \nu\alpha\beta}\,\xi^{a}n_{a}^{\ \beta}k^{\nu}k^{\alpha}$.
Furthermore, using the equation of parallel transport for $n_a^{\ \mu}$ and $v_O^\mu$, eq.~(\ref{parallel_transport}), and projecting the Sachs equation along the spatial basis $n_a^{\ \mu}$, one obtains an evolution equation for $\xi^a$,
\begin{equation}
\label{xibevol}
\frac{\dd^2 \xi^{a} }{\dd \lambda^2}=\mR^a_{\; \; b}\xi^{b}\;,
\end{equation}
where the 2D tensor $\mR_{ab}$ is defined by
\begin{equation}
\mR_{ab}\equiv R_{\mu\nu\rho\sigma}k^{\nu}k^{\rho}n_{a}^{\ \mu}n_{b}^{\ \sigma}\;.
\label{R_def}
\end{equation}

As eq.~(\ref{xibevol}) is linear with initial condition $\xi^a(0)=0$, its solution can be written in the form
\be
\label{eq:xiaxib}
\xi^a=\mD_{ab} \theta_O^b \;,
\ee
where 
\be
\label{angle_O}
 \theta_O^b \equiv \left. \frac{\dd \xi^b}{\dd \lambda}\right|_{\lambda =0}\;
\ee
is the (vectorial) angle between the photon geodesic and the neighboring one at the observer.
$\mD_{ab}$ is a linear matrix, called {\em Jacobi mapping}. It relates the angle of observation $ \theta_O^b$ to the image  on the screen adapted 
to $v_O^\mu$ and $k^\mu$ described by the two spatial components of $\xi^\mu$, $\xi^a$. 
From eq.~(\ref{eq:xiaxib}), eq.~(\ref{xibevol}) can be rewritten as an evolution equation for $\mD_{ab}$ \cite{SEF,1994CQGra..11.2345S,Lewis:2006fu,Uzan:2000xv},
\begin{equation}
\label{Devol}
\frac{\dd^2}{\dd \lambda^2} \mD_{a b}=\mR_{a c}\ \mD_{cb}\;.
\end{equation}
Equation (\ref{angle_O}) and $\xi^a(0)=0$ imply that $\mD_{ab} (0) = 0$ and $\dd \mD_{ab} / \dd \lambda |_{\lambda =0} = \delta_{ab}$. Note that $\mR_{ab}$ is symmetric but $\mD_{ab}$ is generally not.

We can decompose the linear mapping $\mD_{ab}$ into a spin-0 component and a spin-2 component, respectively,
\bea
{}_0 \mD &\equiv& \mD_{11} + \mD_{22} + \ii \left(\mD_{12}-\mD_{21}\right) \;, \\
{}_2 \mD &\equiv &\mD_{11} - \mD_{22} + \ii \left(\mD_{12}+\mD_{21}\right)\;.
\eea
The spin-2 part, ${}_2 \mD$, is directly related to the usual shear spin-2 field $\gamma \equiv \gamma_1 + \ii \gamma_2$. Indeed,
by defining the vectorial angular position at the source as $\theta_S^a \equiv \xi^a / \lambda_S$  one finds the following relations \cite{Uzan:2000xv}
\be
\gamma_1 = -\frac1{2 \lambda_S} (\mD_{11} - \mD_{22})\;, \qquad \gamma_2 = -\frac1{2 \lambda_S} (\mD_{12} + \mD_{21})\;. \label{shear_def}
\ee
The spin-0 part, ${}_0 \mD$, contains a real part, the trace $ \mD= \mD_{11} + \mD_{22}$, which is related to the usual convergence by
\be
 \kappa = 1 -\frac1{2 \lambda_S}  \mD \;.
\label{conv_def}
\ee
The imaginary part comes from the fact that, unlike $\mR_{ab}$, $\mD_{ab}$  is not necessarily symmetric. Indeed, as we will see, it is not symmetric at second order. The imaginary part of ${}_0 \mD$ corresponds to a rotation of the observed object and it is related to the usual rotation parameter $\omega$ by
\be
\omega = -\frac1{2 \lambda_S} (\mD_{12} - \mD_{21})\;.
\ee 
At second order, the rotation has no observational consequences on the observed galaxy polarization.\footnote{More precisely, 
one can define the observed complex shape polarization as $p=(m_{11}-m_{22}+2\ii m_{12})/(m_{11}+m_{22})$ where $m_{ij}$ is the luminosity distribution matrix of observed galaxies. For an unpolarized source and in absence of rotation this is $p=2 g/(1+g g*)$~\cite{Bartelmann:1999yn}. In case of rotation this relation becomes $(1-\tilde\kappa^{*})/(1-\tilde\kappa)\,p=2 \tilde g/(1+\tilde{g} {\tilde g}^*)$ where $\tilde \kappa$ is complex and defined as $\tilde \kappa \equiv \kappa+\ii \omega$ and $\tilde g\equiv g/(1-\tilde\kappa)$. Thus, as $\tilde \kappa$ is real at linear order and $\gamma$ vanishing at zeroth order, the imaginary part of $\tilde \kappa$ enters in the expression of the observed polarization at third order only.} Therefore we will ignore it in the following. We are thus left with one scalar degree of freedom describing the convergence and 2 degrees of freedom for the shear. The latter can be mapped into the so-called ``electric'' and ``magnetic''
modes (see Appendix~\ref{app:spart}). Note that at second order the electric mode is not necessarily equal to the convergence field as in the linear case.

What we observe is the ratio between the anisotropic and isotropic deformations, 
i.e.~the reduced shear, defined as \cite{Bartelmann:1999yn}
\be
g \equiv \frac{\gamma}{1-\kappa}\;.
\ee
From eqs.~(\ref{shear_def}) and (\ref{conv_def}) this is given, in terms of the Jacobi mapping $\mD_{ab}$, by
\be
g = - {{}_2 \mD}/{\mD}\;.
\label{reduced}
\ee

In the following we will solve eq.~(\ref{Devol}) in a perturbed universe. For the background we will assume a 
flat Friedmann-Lema\^{i}tre-Robertson-Walker (FLRW) metric given by $\dd s^2 = a^2(\eta) (-\dd \eta^2 + \dd \bx^2 )$, where $\eta$ is the conformal time. Since null geodesics are not affected by conformal transformations, it will be convenient to perform the calculation without the Friedmann expansion and reintroduce the effect of the expansion only at the end.
Indeed, as shown in Appendix \ref{app:expansion}, the effect of the expansion can be simply taken into account by rescaling the mapping $\mD_{ab}$ by the scale factor, e.g. $\mD_{ab} \to a \mD_{ab}$. Note that, as both ${}_2 \mD$ and $\mD$ get rescaled by the conformal transformation, the reduced shear (\ref{reduced}) is not affected by 
the expansion. We parametrize the photon geodesic such that $x^0(\lambda(\eta) ) =  \eta_0-\eta$.
It is thus convenient to define $\chi\equiv \eta_0-\eta$ so that
\begin{equation}
k^{0}=\frac{\dd \chi}{\dd \lambda}
\end{equation}
and the evolution equation (\ref{Devol}) now reads 
\begin{equation}
\label{D_evol}
\frac{\dd^2}{\dd\chi^2}\mD_{ab} + \frac1{k^0} \frac{\dd k^0}{\dd \chi} \frac{\dd}{\dd \chi} \mD_{ab}= \frac1{(k^0)^2}\mR_{ac} 
\mD_{cb} \;.
\end{equation}
We can solve this second-order differential equation order by order in the metric perturbations.

The solution of eq.~(\ref{D_evol}) gives the linear mapping between the observed angle and 
the shape of a source at a given coordinate time $\eta_S$. Observationally, we are interested in a mapping 
where the source is defined at a given redshift $z_S$. As in an inhomogeneous universe the redshift is a perturbed
quantity,  at second order we expect a contribution to the reduced shear due to the coupling between this perturbation and 
the lenses. Thus, the reduced shear at constant redshift is given by
\be
g_z = g - \frac{\dd g}{\dd z_S} \delta z_S\;,
\label{redshift_correction}
\ee
where $\delta z = z +1- a_0/a$ is the perturbation of the redshift. As this correction is not conformally invariant, it introduces a dependence on the expansion. At first order, though, only the trace of the linear mapping, $\mD$, is affected by redshift perturbations~\cite{2008PhRvD..78l3530B}, but the traceless part is not and therefore in this case $g_z=g$.

%%%%%%%%%%%%%%%%%%%%%%%%%%%%%%%%%%%%%%%%%%%%%%%%%%%%%%%%%%%%
%%%%%%%%%%%%%%%%%%%%%%%%%%%%%%%%%%%%%%%%%%%%%%%%%%%%%%%%%%%%
%%%%%%%%%%%%%%%%%%%%%%%%%%%%%%%%%%%%%%%%%%%%%%%%%%%%%%%%%%%%

\section{The shear and the convergence at first order}
\label{sec:linear}
%%%%%%%%%%%%%%%%%%%%%%%%%%%%%%%%%%%%%%%%%%%%%%%%%%%%%%%%%%%%
%%%%%%%%%%%%%%%%%%%%%%%%%%%%%%%%%%%%%%%%%%%%%%%%%%%%%%%%%%%%
%%%%%%%%%%%%%%%%%%%%%%%%%%%%%%%%%%%%%%%%%%%%%%%%%%%%%%%%%%%%

As a warm up exercise, before the calculation of the shear at second order we derive here, using 
eq.~(\ref{D_evol}), the shear and the convergence at first order.  We consider a perturbed FLRW metric in Newtonian gauge, written in Cartesian coordinates as
\be
\dd s^2= a^2(\eta) \left[-\big(1+2\phi\big)\dd \eta^2+ \big(1-2\psi \big)\dd \bx^2 \right]  ~,
\label{metric1_exp}
\ee
where we have neglected primordial vector and tensor perturbations. For convenience,  we define also the Weyl potential $\Psi$  as 
\be
\Psi \equiv (\phi + \psi)/2 \;,
\ee
and we will use it in the following whenever the combination $\phi+\psi$ appears. As explained in the previous section, the reduced shear (\ref{reduced}) is not affected by the expansion and we can set the scale factor $a=1$.

Let us define $k^\mu(0)\equiv(1,\ve_r)$ as the photon 4-momentum at the observer, where $\ve_r$ defines the direction of the line of sight. 
Note that in this definition we have set the metric perturbations at the observer position to zero. The final result will be independent of this choice. Indeed, since  metric perturbations at the observer do not depend on the direction of observation, they can be reabsorbed into the homogeneous mapping. 
As in flat spacetime the Christoffell symbols vanish, $k^\mu(0)$ is parallel propagated 
along the background geodesic, while the curvature tensor $R_{\mu \alpha \beta \nu}$ vanishes thus making $\mR_{ab}$ at least a first-order quantity.
Then, at first order in perturbations eq.~(\ref{D_evol}) simplifies to
\be
\frac{\dd^2}{\dd\chi^2}\mD_{ab} +  \frac{\dd k^0}{\dd \chi} \frac{\dd}{\dd \chi} \mD_{ab} = \mR_{ac}  \mD_{cb} \;,
\label{Sachs_first}
\ee
where we have used that $\dd k^0 /\dd \chi$ vanishes on the background.

On the background this equation becomes ${\dd^2 \mD_{ab}}/{\dd\chi^2}=0$. Requiring that the homogeneous mapping is proportional to the unit matrix yields $\mD_{ab} = \chi \delta_{ab}$ for the background solution. Furthermore, we can plug the background solution for $\mD_{ab}$  in the second term on the left-hand side and on the right-hand side of eq.~(\ref{Sachs_first}) to obtain
\be
\frac{\dd^2}{\dd\chi^2}\mD_{ab} +  \frac{\dd k^0}{\dd \chi} \delta_{ab} = \chi \mR_{ab} \;.
\label{Sachs_first2}
\ee
We then integrate this equation once up to the source using the boundary condition $\dd \mD_{ab} /\dd \chi |_O =\delta_{ab}$,
\be
\frac{\dd}{\dd \chi} \mD_{ab} = (2-k^0) \delta_{ab} + \int_0^\chi \dd \chi' \chi'\; \mR_{ab}\;.
\label{1st_1st}
\ee
The solution of this equation, after an integration by parts, can be written as
\be
\mD_{ab} (\chi_S) = \int_0^{\chi_S} \dd \chi (2-k^0)\delta_{ab} + \int_0^{\chi_S} \dd \chi (\chi_S-\chi) \chi \mR_{ab} \;;
\label{map_first_order}
\ee 
$k^0$ can be obtained by solving the photon geodesic (\ref{geodesic}) at first order, 
\be
k^0=1 -2\phi+\int_0^\chi d\chi' 2 \dot \Psi  \;,
\label{k^0}
\ee
where the dot denotes a partial derivative with respect to $\chi$, i.e.~$\dot{} \equiv \partial / \partial \chi$.  

Let us now compute $\mR_{ab}$. We denote by $n_a^{\ \mu}(0) \equiv (0, \ve_a)$ the spatial basis at the observer. As for the 4-momentum of the photon, $n_a^{\ \mu}(0)$ is parallel propagated along the background geodesic.
Thus, according to the definition (\ref{R_def}), in order to compute $\mR_{ab}$ we need to contract the curvature tensor at first order with the unperturbed 
$k^\mu(0)$ and $n_a^{\ \mu}(0)$, which yields
\be
\mR_{ab}=- \en_a^{\ i} \en_b^{\ j}   2 \Psi_{,ij}  -\frac{\dd^2\psi}{\dd\chi^2} \delta_{ab}\;.
\label{R_first_order}
\ee
Plugging $\mR_{ab}$ given in this equation and the expression for $k^0$ given in eq.~(\ref{k^0}) into eq.~(\ref{map_first_order}) and integrating by parts $\dd^2 \psi/\dd \chi^2$ we obtain the Jacobi mapping at linear order,
\be
\mD_{ab}(\chi_S) = (1 - \psi (\chi_S) ) \chi_S \delta_{ab}+  \int_0^{\chi_S} \dd \chi 
\left[ 4 \Psi -  2  (\chi_S - \chi) \dot \Psi \right] \delta_{ab} 
-  \en_a^{\ i} \en_b^{\ j} \int_0^{\chi_S} \dd \chi (\chi_S - \chi) \chi   \; 2 \Psi_{,ij} \;.
\label{linear_mapping}
\ee
According to the Born approximation we can evaluate the integral along the background geodesic so that 
in this expression $\Psi = \Psi(\chi, \ve_r \chi) $.

Note that the metric (\ref{metric1_exp}) is conformal to 
$\dd s^2 =- (1+4 \Psi) \dd \eta^2 + \dd \bx^2  $.  
Thus, as photon geodesics are conformally invariant, one would naively expect $\mD_{ab}$ to
depend only on the combination $\Psi=(\phi+\psi)/2$ \cite{Lewis:2006fu}. This is too quick a conclusion.
Indeed, parallel transport of the basis $n_a^{\ \mu}$ is not conformally invariant and the 
basis is deformed by the spatial curvature at the source position. 
For this reason the first term of eq.~(\ref{linear_mapping}) also depends on the curvature potential at the source, $\psi_S$.

We now want to extract the shear and convergence from the mapping $\mD_{ab}(\chi_S)$. 
In order to do so, it is useful to introduce spin operators on the sphere 
(see Appendix~\ref{app:spart}). To each point of a 2D Riemannian manifold described by an orthonormal basis $\{\ve_1, \ve_2\}$ we can associate 
a spin-\textit{s} field ${}_sX$ such that under the rotation of $\ve_a$ by an angle $\alpha$ 
it transforms as ${}_sX \to e^{\ii \alpha s} {}_sX$ 
(for more details see \cite{Goldberg1967,Lewis:2001hp,NewmanPenrose66}). 
More precisely, following  \cite{Lewis:2001hp}, the local freedom in the choice of the basis is equivalent to the transformations
\be
\ve_\pm\equiv\ve_1\pm \ii \ve_2\to e^{\ii \alpha} \ve_\pm\; .
\ee
To every spin-\textit{s} ${}_s X$ we can associate a 
symmetric and trace-free tensor of rank $s\ge 0$, $X_{a_1 \ldots a_s}$:
for $s \ge 0$,
\be
X^{a_1\ldots a_s} \equiv 2^{-s} {}_s X  \en_-^{a_1} \cdots \en_-^{a_s}\;.
\ee
The inverse relation is
\be
\label{xs}
{}_s X \equiv  \en_+^{a_1} \cdots \en_+^{a_s} X_{a_1 \ldots a_s}\;.
\ee
For $s<0$ we define $X^{a_1 \cdots a_{|s|}} \equiv 2^{-|s|} {}_s X  \en_+^{a_1} \cdots \en_+^{a_{|s|}}$.

As we are interested in describing the lensing field on the sphere of the sky, for the orthonormal basis we choose the two coordinate basis vectors of a spherical polar coordinate system, $\{\ve_1, \ve_2\}\equiv\{\ve_\theta, \ve_\varphi\}$. Let us rewrite the three spatial vectors $\ve_r$ and $\ve_a$ in the Cartesian representation as
\begin{eqnarray}
\ve_r &=& (\sin \theta \cos \varphi , \sin \theta \sin \varphi, \cos \theta)\;, \label{e_r} \\
\ve_\theta&=& (\cos \theta \cos \varphi , \cos \theta \sin \varphi, -\sin \theta)\;, \\
\ve_\varphi &=& (-\sin \varphi , \cos \varphi, 0) \label{e_varphi}\;,
\end{eqnarray}
where $\theta,\varphi$ are the angles of observation. Furthermore, we can define operators that increase or decrease the index of the spin by 1,
\be
\spart \; {}_s X   \equiv - \sin^s \theta (\pa_\theta + \ii \csc \theta \pa_\varphi) 
(\sin^{-s} \theta)
\; {}_s X \;, 
\qquad \spartb \; {}_s X   \equiv - \sin^{-s} \theta (\pa_\theta - \ii \csc \theta \pa_\varphi) (\sin^{s} \theta) \; {}_s X \;.
\ee
With the definitions (\ref{e_r})--(\ref{e_varphi}) above we have
\be
e_r^i \pa_i=\pa_r, \qquad \en_\theta^i\pa_i=\frac{1}{\chi}\pa_\theta, \qquad \en_\varphi^i\pa_i=\frac{1}{\chi\sin\theta}\pa_\varphi\;,
\ee
where we have used that $r=\chi$ along the photon geodesic.
Using these relations it is easy to verify that, if $X=X(\chi)$ is a scalar field,
\be
  \en_+^{\ i} X_{,i} =- \frac1\chi \spart X \;, \qquad 
 \en_-^{\ i} X_{,i} = - \frac1\chi \spartb X\;.
\label{n+}
\ee
Then, employing the useful relation $\chi \en_\pm^{\ i} \partial_i \en_\pm^{\ j} = \cot \theta \en_\pm^{\ j}$ one can verify that
\be
\en_+^{\ i}\en_+^{\ j} X_{,ij} = \frac1{\chi^2} \spart^2 X   \;, \qquad
\en_-^{\ i} \en_-^{\ j} X_{,ij} = \frac1{\chi^2} \spartb^2 X  \;,
\label{n+n+}
\ee
and, analogously with $\chi \en_\mp^{\ i} \partial_i \en_\pm^{\ j} = - \cot \theta \en_\pm^{\ j} - 2 \ek^{\ j}$, that
\be
\en_+^{\ i}\en_-^{\ j} X_{,ij} = 
\en_-^{\ i} \en_+^{\ j} X_{,ij} = \frac1{\chi^2}  \spartb \spart  X - \frac2\chi X_{,r}   \;.
\label{n+n-}
\ee
Note that the 2D Laplacian on the sphere is given by $\spart \spartb X = 
\spartb \spart X$.

Let us first apply these definitions to $\mR_{ab}$. By making use of eq.~(\ref{xs}) 
we can define a spin-2 field on the sphere, ${}_{2}\mR\equiv \en_+^{\ a}\en_+^{\ b}\mR_{ab}$, which by eq.~(\ref{n+n+}) reads
\be
{}_2\mR = - \frac{2}{\chi^2} \spart^2 \Psi\;.
\label{2R}
\ee
Then, using that $\en_a^{\ i} \en_a^{\ j}
= (\en_+^{\ i} \en_-^{\ j}+ \en_-^{\ i} \en_+^{\ j})/2 $ and eq.~(\ref{n+n-}), the trace of $\mR_{ab}$ reads
\be
\mR = - \frac4\chi \Psi_{,r} - \frac{2}{\chi^2} \spartb \spart \Psi - 2 \frac{\dd^2 \psi }{\dd \chi^2}  \;.
\label{traceR}
\ee
We can do the same for the mapping (\ref{linear_mapping}). One finds the  spin-2 mapping field,
\be
{}_2 \mD(\chi_S) = -  2 \int_0^{\chi_S} \dd \chi \frac{\chi_S -\chi}{\chi} \;   
\spart^2 \Psi \;.
\label{2D}
\ee
At first order the reduced shear $g$ is given by ${}_2 \mD$ divided by the 
background part of $\mD$, which is simply $\bar \mD = 2 \chi_S$, i.e.
\be
g =  \int_0^{\chi_S} \dd \chi \frac{\chi_S -\chi}{\chi \chi_S} \;   
\spart^2 \Psi \;.
\label{shear_first}
\ee
Taking the trace of eq.~(\ref{linear_mapping}) and defining $\delta \mD = \mD - 2 \chi$ one finds, after an integration by part, 
\be
\delta \mD(\chi_S) = - 2 \psi(\chi_S) \chi_S + 2 \int_0^{\chi_S} \dd \chi \left( 2 \Psi  - \frac{\chi_S -\chi}{\chi}   { \spartb}  
\spart \Psi \right)\;,
\label{traceD}
\ee
which is proportional to the convergence, $\kappa$.
The first two terms on the right-hand side 
are not usually included in the convergence because they are negligible on small angular scales.\footnote{The Laplacian $\spartb \spart$ introduces a factor $l (l+1)$ in harmonic space, thus enhancing the contribution to the power spectrum from the third term of eq.~(\ref{traceD}) with respect to the contribution of the other terms.} 
In particular, the first term 
is a relativistic effect due to the deformation of the size of the source
caused by the curvature potential at the source. The second term is the Shapiro time delay \cite{Shapiro:1964uw}. 
Both effects are negligible on small angular scales.

We can check that the first two terms in eq.~(\ref{traceD}) contribute to the convergence by considering a nonexpanding FLRW universe with nonzero constant spatial curvature $K$ and metric
\be
\label{fbmetdef1}
ds^2=-\dd\eta^2+\frac{1}{\big(1+\psi_K (r) \big)^2} (dr^2+r^2 d\Omega^2) \qquad \mbox{with} \qquad \psi_K (r) \equiv \frac{K r^2}{4}\;.
\ee
In this universe the angular diameter distance is
\be
\label{dconv}
D_A(\chi)=\left\{ \begin{array}{ll} {\sin (\sqrt{K}\chi) }/{\sqrt{K}} &\quad \hbox{for}\quad K>0\; ,\\
 {\sinh (\sqrt{|K|}\chi )}/{\sqrt{|K|}} &\quad  \hbox{for} \quad K<0\; .
\end{array} \right.
\ee
For small curvature, $K \ll 1$, the expression above can be expanded at first order in $K$ to give
\be
\label{dAK}
D_A(\chi_S)=\chi_S-\frac{K\chi_S^3}{6}\; .
\ee
The first term on the right-hand side is the angular diameter distance in a spatially flat universe while the second term is a small perturbation to it, due to the curvature. 

As the trace ${\cal D}$ corresponds to twice the angular diameter distance, we can check that $\delta {\cal D}(\chi_S)/2$, with $\delta {\cal D}(\chi_S)$ given in eq.~(\ref{traceD}), correctly reproduces this small perturbation in this particular case. For $\phi=0$ and $\psi =\psi_K$ (and $a=1$) the flat perturbed metric in Poisson gauge, eq.~(\ref{metric1_exp}), reproduces at first order in $\psi_K$ the spatially curved metric (\ref{fbmetdef1}). Thus, in this case $\Psi \equiv (\phi+\psi)/2 = {K r^2}/{8}$. As the Weyl potential $\Psi$ depends only on the radial coordinate, the last term on the right-hand side of eq.~(\ref{traceD}) vanishes so that it does not contribute to the angular diameter distance. Furthermore, the contributions from the first and the second term of eq.~(\ref{traceD}) are, respectively, $-{K\chi_S^3}/{4} $ and ${K\chi_S^3}/{12}$, where we have used that, at lowest order in $\psi_K$, $r=\chi$ along the photon geodesic. This yields $\delta {\cal D}(\chi_S)/2 = - {K\chi_S^3}/{6}$, i.e.~the second term of eq.~(\ref{dAK}).
Note that the first two terms of eq.~(\ref{traceD}) have no counterpart in the shear. Indeed, a constant curvature deviates light rays only isotropically. Furthermore, Kaiser's relation \cite{1995ApJ...439L...1K} between shear and convergence,  $ \spart\kappa=\spartb \gamma$, is only valid in the limit of small angles, when the first two terms of eq.~(\ref{traceD}) are negligible.

%%%%%%%%%%%%%%%%%%%%%%%%%%%%%%%%%%%%%%%%%%%%%%%%%%%%%%%%%%%%%%%%%%%%%%%%%%%%%%%%%%%%%%%%%%%%%%%%%%%%%%%%%%%%%%%%
%%%%%%%%%%%%%%%%%%%%%%%%%%%%%%%%%%%%%%%%%%%%%%%%%%%%%%%%%%%%%%%%%%%%%%%%%%%%%%%%%%%%%%%%%%%%%%%%%%%%%%%%%%%%%%%%
%%%%%%%%%%%%%%%%%%%%%%%%%%%%%%%%%%%%%%%%%%%%%%%%%%%%%%%%%%%%%%%%%%%%%%%%%%%%%%%%%%%%%%%%%%%%%%%%%%%%%%%%%%%%%%%%

\section{The shear at second order}
\label{sec:shear_2nd}

%%%%%%%%%%%%%%%%%%%%%%%%%%%%%%%%%%%%%%%%%%%%%%%%%%%%%%%%%%%%%%%%%%%%%%%%%%%%%%%%%%%%%%%%%%%%%%%%%%%%%%%%%%%%%%%%
%%%%%%%%%%%%%%%%%%%%%%%%%%%%%%%%%%%%%%%%%%%%%%%%%%%%%%%%%%%%%%%%%%%%%%%%%%%%%%%%%%%%%%%%%%%%%%%%%%%%%%%%%%%%%%%%
%%%%%%%%%%%%%%%%%%%%%%%%%%%%%%%%%%%%%%%%%%%%%%%%%%%%%%%%%%%%%%%%%%%%%%%%%%%%%%%%%%%%%%%%%%%%%%%%%%%%%%%%%%%%%%%%

In this section we will compute the lensing shear at second order. As we concentrate on the shear, we will compute only the trace-free and symmetric part of the matrix $\mD_{ab}$. 
Indeed, at second order the trace and the antisymmetric part contribute only to the convergence $\kappa$ and the rotation $\omega$, respectively.
We will do the calculation in the
so-called generalized Poisson gauge, where the second-order metric can be written as \cite{Bruni:1996im}
\be
\label{metric2}
\dd s^2=a^2(\eta) \left[ -e^{2\phi}\dd \eta^2+2 \omega_{i}\,\dd \eta\dd x^{i}
+ \left(e^{-2\psi} \delta_{ij} +h_{ij} \right) \dd x^{i}\dd x^{j}\right]\;.
\ee
Here the vector component $\omega_i$ is divergenceless, $\partial_i \omega_i=0$, and the tensor component $h_{ij}$ 
is divergenceless and traceless, $\partial_i h_{ij} = 0 = h_{ii}$. As we neglect primordial vector and 
tensor perturbations, $\omega_i$ and $h_{ij}$ are only second-order quantities. Note that we have used the exponential
form for the gravitational potentials in the metric. Indeed, in this form the metric is conformal to $\dd s^2 = -e^{4\Psi} \dd \eta^2 +2 \omega_{i}\,\dd \eta\dd x^{i}
+ \left(\delta_{ij} +h_{ij} \right) \dd x^{i}\dd x^{j} $ so that the effect of scalar perturbations on the 
null geodesic is only through the Weyl potential $\Psi = (\phi + \psi)/2$.
As done in the previous section, 
we will drop the effect of the expansion setting 
$a=1$. We will reintroduce the expansion in Sec.~\ref{sec:reduced}.

%%%%%%%%%%%%%%%%%%%%%%%%%%%%%%%%%%%%%%%%%%%%%%%%%%%%%%%%%%%%
\subsection{Solving the Sachs equation}
%%%%%%%%%%%%%%%%%%%%%%%%%%%%%%%%%%%%%%%%%%%%%%%%%%%%%%%%%%%%

To solve the Sachs equation at second order let us go back to 
eq.~(\ref{D_evol}). This can be integrated to obtain
\be
\frac{\dd}{\dd \chi} \mD_{ab} = \delta_{ab} + 
\int_0^\chi \dd \chi' \left( \frac{1}{(k^0)^2} \mR_{ac} \mD_{cb} 
- \frac1{k^0} \frac{\dd k^0}{\dd \chi} \frac{\dd}{\dd \chi} \mD_{ab} \right)\;,
\label{first_integration}
\ee
where we have used that $\dd \mD_{ab} /\dd \chi |_O=\delta_{ab}$ at the observer. 
Since $\dd k^0 /\dd \chi$ vanishes on the background we can rewrite 
the last term of this equation using the first-order equation (\ref{1st_1st})
and integrate eq.~(\ref{first_integration}) up to the source $S$. We obtain, after integration by parts,
\be
\mD_{ab}(\chi_S) = \chi_S \delta_{ab} + \int_0^{\chi_S} \dd \chi \frac{\chi_S - \chi}{\chi} \mS_{ab}\;,
\label{D_S}
\ee
where we have defined the source term $\mS_{ab}$ as
\be
\mS_{ab} \equiv  \frac{\chi}{(k^0)^2} \mR_{ac} \mD_{cb} 
- \frac{\chi}{k^0} \frac{\dd k^0}{\dd \chi} (2-k^0) \delta_{ab} - \chi \frac{\dd k^0}{\dd \chi} \int_0^\chi \dd \chi' \chi' \mR_{ab}\;.
\label{source_inter}
\ee

At second order we need to go beyond the standard Born approximation. Since the 
source term $\mS_{ab}$ is at least first order, we just need to integrate along a geodesic which is 
perturbed at first order. Thus, we evaluate the source term at the perturbed geodesic position 
$x^i_{\rm pert}(\chi) = x^i(\chi) + \delta x^i(\chi)$, where $\delta x^i$ is the geodesic deviation.
Expanding $\mS_{ab}$ along the background geodesic,
\be
\mS_{ab} (x^i_{\rm pert}) = \mS_{ab}(x^i) + \delta x^j \cdot \delta(\mS_{ab})_j |_x \;,
\ee
where the shift $\delta(\mS_{ab})_j$ will be computed below,
we can rewrite the source term $\mS_{ab}$ as a function of the unperturbed geodesic position $x^i$ as
\be
\mS_{ab} \equiv  \frac{\chi^2}{(k^0)^2} \mR_{ab} + \chi \mR_{ac} \; \delta \mD_{cb} 
- \frac{\chi}{k^0} \frac{\dd k^0}{\dd \chi} (2-k^0) \delta_{ab} - \chi \frac{\dd k^0}{\dd \chi} \int_0^\chi \dd \chi' \chi' \mR_{ab} + \chi^2 \delta x^j \cdot \delta(\mR_{ab})_j\;,
\label{source_term}
\ee
where to write the first two terms 
we have separated $\mD_{ab}$ into its background and first-order part as $\mD_{ab} = \chi \delta_{ab} + \delta \mD_{ab}$ and we have employed for the last term that at leading order $\mS_{ab} = \chi^2 \mR_{ab}$. We will now compute the symmetric traceless part 
of the source $\mS_{ab}$ evaluating one by one each of the terms on the right-hand side of eq.~(\ref{source_term}).

Let us start with the first term in eq.~(\ref{source_term}). Since on a flat background spacetime 
the unperturbed part of $R_{\mu \alpha \beta \nu}$ vanishes, according to the definition (\ref{R_def}), i.e., $\mR_{ab}=R_{\mu \alpha \beta \nu} k^\alpha k^\beta n_a^{\ \mu} n_b^{\ \nu}$,
in order to compute $\mR_{ab}$ up to second order we just need to consider the 
photon wave vector $k^\mu$ and the basis $n_a^{\  \mu}$ up to first order. Integrating the
geodesic equation (\ref{geodesic}) one obtains, up to first order,
\bea
k^0&=&1 -2\phi+\int_0^\chi d\chi' \; 2\dot{\Psi} \;, \\
k^i&=&( 1 +2\psi )\ek^{\ i}-\int_0^\chi d\chi' \; 2 \pa^i \Psi\;.
\eea
Integrating up to first order the parallel transport equation for $n_a^{\  \mu}$, eq.~(\ref{parallel_transport}), 
one finds
\bea
n_a^{\ 0}&=&- \en_a^{\ i} \int_0^\chi d\chi' \pa_i\phi \;, \\
n_a^{\ i}&=& \en_a^{\ i} \psi + \ek^{\ i} \en_a^{\ j} \int_0^\chi d\chi' \pa_j\psi\;.
\eea
Since  
the vector and tensor parts of the metric are already second order, they enter linearly in $R_{\mu \alpha \beta \nu}$, which just needs to be contracted with the unperturbed $k^\mu$ and $n_a^{\ \mu}$.  
Furthermore, it is convenient to compute the ratio $k^i/k^0$ which appears when combining $k^i$ in the definition (\ref{R_def})
with the denominator of the first term of eq.~(\ref{source_term}). This yields
\be
\frac{k^i}{k^0}= \ek^{\ i} (1+ 2 \Psi) -  \en_a^{\ i} \en_a^{\ j} \int_0^\chi \dd \chi' \; 2 \Psi_{,j}\;.
\label{ratiok}
\ee

Finally, computing $R_{\mu \alpha \beta \nu}$ up to second order and making use of the definition (\ref{R_def}),
the traceless part of the first term of eq.~(\ref{source_term}) reads
\be
\begin{split} 
 \frac{\chi^2}{(k^0)^2}\left( \mR_{ab} \right)^T =& - \chi^2 \;  \en_{(a}^{\ i} \en_{b)}^{\ j} \left( 2 \Psi_{,ij}+  4 \Psi_{,i} \Psi_{,j} + 8 \Psi \Psi_{,ij} 
+ 8  \ek^{\ k} \Psi_{,ik} \int_0^\chi \Psi_{,j} \right. \\
& \left. -  \ek^{\ k} \omega_{k,ij} + \frac{\dd}{\dd \chi} \omega_{i,j} - \frac12 \ek^{\ k} \ek^{\ l}  h_{kl,ij} 
- \frac12 \frac{\dd^2}{\dd \chi^2} h_{ij} + \frac12 \frac{\dd}{\dd \chi} \ek^{\ k} h_{k i,j}\right) \;,
\end{split}
\ee
where the parenthesis in the indices denote symmetrization.
After few manipulations, employing eqs.~(\ref{n+}) and (\ref{n+n+}), and the useful relation $\chi \en_+^{\ i} \partial_i \ek^{\ j} = \en_+^{\ j}$, we can write its 
contribution to ${}_2\mS \equiv  \en_+^{\ a } \en_+^{\ b } \mS_{ab}$ as
\be
\begin{split}
%{}_2\mS \supset
&  -  2 \spart^2 \Psi - 4 \spart \Psi \spart \Psi - 8 \Psi \spart^2 \Psi 
- 8 \left( \chi \spart \Psi_{,r} - \spart \Psi \right) \int^\chi_0 \dd \chi' \frac1{\chi'} \spart \Psi \\
&  +  \spart^2 \omega_r 
+ \frac{\dd}{\dd  \chi}(\chi \spart \; {}_1 \omega) + \frac12\spart^2 h_{rr} 
+ \frac{\dd}{\dd \chi}(\chi \spart \; {}_1h_r) + \frac{\chi}2 \frac{\dd^2}{\dd \chi^2} (\chi \; {}_2h)  \;,
\label{term1}
\end{split}
\ee
where $\omega_r \equiv \ek^{\ i} \omega_i$,  $h_{rr} \equiv \ek^{\ i}\ek^{\ j} h_{ij}$ and we have defined the spin-1 part of $\omega$, ${}_1 \omega 
\equiv \en_{+}^{ \ i} \omega_i$, the spin-1 quantity ${}_1h_r \equiv \en_+^{ \ i} \ek^{\ j} h_{ij} $ and 
the spin-2 part of the tensor mode $h_{ij}$, ${}_2h \equiv \en_+^{ \ i} \en_+^{ \ j} h_{ij}$.\footnote{Note that  $\omega_i$ and $h_{ij}$ can be decomposed respectively as $\omega_i = \omega_r \; \en_r^{\ i} + \frac12 {}_{-1}\omega \; \en_{+}^{\ i}+ \frac12 {}_1\omega\;  \en_{-}^{\ i} $ and $h_{ij} =h_{rr}\; \big( \en_r^{\ i} \en_r^{\ j}-\frac12 \en_{+}^{(i}\en_{-}^{j)} \big)  + {}_{-1} h_r \; \en_{+}^{\ (i} \en_r^{\ j)} + {}_1 h_r\; \en_{-}^{\ (i} \en_r^{\ j)}  + \frac14{}_{-2} h\;  \en_{+}^{\ i}\en_{+}^{\ j} +\frac14 { }_2 h\; \en_{-}^{\ i}  \en_{-}^{\ j} $. These components are not independent, as they are related by the divergenceless  conditions $\omega_{i,i}=0=h_{ij,i}$.}

The second term of eq.~(\ref{source_term}) is the product of two first-order quantities, $\mR_{ac}$ and $\delta \mD_{cb}$, 
both symmetric at first order and containing 
a trace and a traceless part. The symmetric traceless part of this product is simply given by
\be
\chi (\mR_{ac} \delta \mD_{cb})^{ST} =  \chi (\mR_{ab})^T \frac12 \delta \mD + \chi \frac12 \mR ( \delta \mD_{ab})^T \;,
\ee
where $\delta \mD$ is the perturbed trace of the mapping $\mD_{ab}$ and $\mR$ 
is the trace of $\mR_{ab}$. Note that the term $\chi\, \mR_{ac}^{T}\, \delta \mD_{cb}^T$ contains a trace and an  antisymmetric part and thus contributes only to the second-order convergence and rotation.
Using the expressions  for ${}_2 \mR$ and $\mR$, eqs.~(\ref{2R}) and (\ref{traceR}),
and those for ${}_2 \mD$ and $\mD$, eqs.~(\ref{2D}) and (\ref{traceD}), one obtains for the contribution of the second term to ${}_2\mS$,
\be
\begin{split}
%{}_2 \mS \supset 
& \; 2 \psi \spart^2 \Psi -\frac2\chi \spart^2 \Psi \int_0^\chi \dd \chi' \left( 
2 \Psi - \frac{\chi-\chi'}{\chi'} \spartb \spart \Psi \right)   \\ 
&  + 2 \left( 2 \Psi_{,r} 
+ \frac{1}{\chi} \spartb \spart \Psi +  \chi \frac{\dd^2 \psi }{\dd \chi^2}  \right) 
\int_0^\chi \dd \chi' \frac{\chi -\chi'}{\chi'} \; \spart^2 \Psi   \;.
\label{term2}
\end{split}
\ee
The first line is $ \chi\;{}_2\mR  \delta \mD$ while the second is $\chi   \mR \; {}_2\delta\mD$.

The third term of $\mS_{ab}$ is a pure trace so that it does not contribute to the shear. 
The traceless part of the fourth term of eq.~(\ref{source_term}) can be straightforwardly 
computed by noting that
\be
\frac{\dd k^0}{\dd \chi} = -2 \frac{\dd \phi}{\dd \chi} + 2 \dot \Psi \;.
\ee 
Thus, its contribution to the source ${}_2\mS$ is
\be
%{}_2 \mS \supset
4 \chi \left( - \frac{\dd \phi}{\dd \chi} +  \dot \Psi \right) \int_0^\chi \dd \chi' \frac1{\chi'} \spart^2 \Psi \;.
\label{term3}
\ee

In order to express the last term of eq.~(\ref{source_term}) in terms of a spin-2 field 
we need to solve the geodesic equation at first order. 
We can solve $\dd x^i/\dd \chi = k^i/k^0$ making use of eq.~(\ref{ratiok}). 
After integrating by parts this yields the geodesic deviation, 
\be
\delta x^i=\ek^{\ i} \int_0^\chi d\chi' \; 2\Psi - \en_a^{\ i} \en_a^{\ j} 
\int_0^\chi d\chi'(\chi-\chi') \;2 \Psi_{,j}\;.
\label{deviation}
\ee
The shift of $\mR_{ab}$, $\delta(\mR_{ab})_i$, is simply 
given by the variation  of $\Psi$ and $\dd^2 \psi/\dd \chi^2$ along the geodesic
in eq.~(\ref{R_first_order}). We only need to take the traceless part of $\delta(\mR_{ab})_i$. Thus we have
\be
\delta(\mR_{ab} )^T_i = - \en_a^{\ j} \en_b^{\ k} \; 2\Psi_{,ijk}\;,
\ee
which can be contracted with $\delta x^i$ of eq.~(\ref{deviation}). Note that, as we are varying 
directly the scalar Weyl potential $\Psi$, it was not necessary to introduce a covariant derivative on the sphere as in \cite{Challinor:2002cd}.
With the definitions and relations of Sec.~\ref{sec:linear} it is possible to verify that
\be
\en_+^{\ i} \en_+^{\ j} \en_+^{\ k} X_{,ijk} = - \frac{1}{\chi^3} \spart^3 X \;, \qquad
\en_+^{\ i} \en_+^{\ j} \en_-^{\ k} X_{,ijk} = - \frac{1}{\chi^3} \spartb \spart^2 X 
- \frac4{\chi^2} \left(\spart X_{,r} - \frac1\chi \spart X \right)  \;. \qquad
\ee
Using these relations and $\en_a^{\ i} \en_a^{\ j} 
= (\en_+^{\ i} \en_-^{\ j}+ \en_-^{\ i} \en_+^{\ j})/2 $ 
one finds, for this last contribution,
\begin{equation}
\begin{split}
& %{}_2 \mS  \supset
4 \left( \frac2{\chi} \spart^2 \Psi - \spart^2 \Psi_{,r} \right) \int_0^\chi \dd \chi' \Psi  \\
 &+ \frac2\chi \spart^3 \Psi  \int_0^\chi \dd \chi' \frac{\chi-\chi'}{\chi'} 
\spartb \Psi + \frac2\chi \spartb \spart^2 \Psi  \int_0^\chi \dd \chi' \frac{\chi-\chi'}{\chi'} 
\spart \Psi + 8 \left(\spart \Psi_{,r} - \frac1\chi \spart \Psi \right) \int_0^\chi  \dd \chi' \frac{\chi-\chi'}{\chi'} \spart \Psi \;, \label{term4}
\end{split}
\end{equation}
where the first and second lines come from contracting with the first  and second terms 
on the right-hand side of eq.~(\ref{deviation}), respectively.
Using the commutation rule for the spin raising and lowering operators \cite{Lewis:2001hp},
\be
(\spartb \spart - \spart \spartb) {}_s X = 2 s \; {}_s X\;,
\ee
it is convenient to rewrite the second term in the second line of eq.~(\ref{term4}) as
\be
\label{term5}
\frac2\chi \spartb \spart^2 \Psi  \int_0^\chi \dd \chi' \frac{\chi-\chi'}{\chi'} \spart \Psi
= \frac2\chi \spart \spartb \spart  \Psi  \int_0^\chi \dd \chi' \frac{\chi-\chi'}{\chi'} \spart \Psi +
\frac4\chi \spart  \Psi  \int_0^\chi \dd \chi' \frac{\chi-\chi'}{\chi'} \spart \Psi \;.
\ee

Finally, combining eqs.~(\ref{term1}), (\ref{term2}), (\ref{term3}), (\ref{term4}) and (\ref{term5}),   
replacing partial derivatives with respect to $r$ by using $\partial_r = \dd/\dd \chi - \partial_\chi$ (we remind the reader that $\dot{\Psi}=\pa_\chi \Psi$)
and integrating the total derivatives by parts, we obtain
\be
\begin{split}
{}_2 \mD(\chi_S)=&\;
2 \int_0^{\chi_s} d\chi \frac{\chi_S -\chi}{\chi}
\spart\bigg[ -\spart\Psi+\frac{1}{\chi}\left(\spart^2 \Psi \int_0^{\chi} d\chi' \frac{\chi-\chi'}{\chi'}  \spartb \Psi  
+\spartb\spart \Psi \int_0^{\chi} d\chi' \frac{\chi-\chi'}{\chi'}  \spart \Psi  \right) \bigg]\\
&+4 \int_0^{\chi_s} d\chi \frac{\chi_S -\chi}{\chi} \Bigg[ \frac12 \spart^2   \Psi^2 
 + \frac12\psi(\chi_S) \spart^2\Psi 
- \Psi\ \frac{1}{\chi} \int_0^\chi d\chi' \spart^2\Psi 
+\spart^2 \left( \dot{\Psi}\int_0^\chi d\chi' \Psi \right)
\\
&+
 \spart\Psi \frac1\chi\int_0^{\chi}d\chi'\frac{\chi-\chi'}{\chi'}\spart\Psi 
 \Bigg] -4 \int_0^{\chi_S} d\chi \Bigg[\Psi\int_0^\chi d\chi' \frac{1}{\chi'}\spart^2\Psi 
+ \frac{1}{\chi}\spart^2 \left( \Psi\int_0^\chi d\chi'\Psi \right)
\Bigg]  \\
&+   \frac{\chi_S}2 {}_2 h(\chi_S) +   \int_0^{\chi_S} d\chi \left[ \frac{\chi_S -\chi}{\chi}  \spart^2 (\omega_r + \frac12 h_{rr})
+ \frac{\chi_S}{\chi}\spart ( {}_1 \omega+  {}_1h_r) 
\right]
\;. \label{shear_tot}
\end{split}
\ee
This is the solution of the Sachs equation. It is written in terms of spin raising and lowering operators $\spart$ and $\spartb$, which are just the extensions on the sphere of the usual angular gradients: for small angles one can replace them by angular gradients $\vec \nabla_{\ek} $ transverse to the line of sight. The advantage of using this representation is that the eigenfunctions of these operators are simply spin-weighted spherical harmonics (see Appendix B). 

In the small-angle approximation, i.e.~when transverse scales are much smaller than radial scales, eq.~(\ref{shear_tot}) reduces to its first line. Indeed,  in harmonics space the presence of the operators $\spart$
and $\spartb$ is associated to an $l$ factor. Thus, when observations are confined to large $l$ one gets a larger contribution from those terms containing more $\spart$ and $\spartb$ operators, such as the first line. The parentheses of the first line contain the usual couplings considered in the literature \cite{1997A&A...322....1B,Cooray:2002mj,2003MNRAS.344..857T,Dodelson:2005zj,2006JCAP...03..007S}, 
i.e.~the lens-lens correction and the correction to the Born approximation.
 Note that there are other couplings at play: the nonlinear growth of dark matter fluctuations induced by gravity on small scales introduces other second-order effects that are incorporated in the Weyl potential $\Psi$.\footnote{To a large extent, these effects are those that are expected to dominate in the current surveys and that lead to detectable effects~\cite{1997A&A...322....1B,2002A&A...389L..28B,2003A&A...397..405B}.}

In the small-angle approximation 
the last three lines of eq.~(\ref{shear_tot}) are suppressed by the ratio between the transverse 
scales probed and the longitudinal distances. However, for a full-sky survey where one probes 
larger scales they can become important. Let us list here these terms collecting them by their 
physical interpretation (for simplicity, integration over $\dd\chi$ and $\dd \chi'$ will be omitted):
\begin{itemize}
\item An {\em intrinsic contribution}, integrated only once along the line of sight,
\be
2 \frac{\chi_S-\chi}{\chi}  \spart^2 \Psi^2 (\chi)\;, \label{new1}
\ee
coming from the Riemann tensor at second order. This is a purely general relativistic effect of 
second-order gravity. 
\item A {\em source-lens} coupling,
\be
2 \frac{\chi_S-\chi}{\chi} \psi(\chi_S)  \spart^2 \Psi^2 (\chi)\;,
\ee
which comes from the coupling between the lens and the curvature at the source, inducing a deformation
of its shape. As we will see, this term is absent from the final expression of the reduced shear, as it cancels with an equivalent term coming from the corrections due to 
the denominator of eq.~(\ref{reduced}).
\item {\em Time delay-lens} couplings,
\be
\begin{split}
& 4\frac{\chi_S -\chi}{\chi} \left( 
- \Psi(\chi) \frac{1}{\chi} \spart^2\Psi(\chi') 
+\spart^2 \dot{\Psi}(\chi)  \Psi(\chi') + \dot{\Psi}(\chi) \spart^2 \Psi (\chi') \right)\\
&-4 \left(\Psi (\chi)  \frac{1}{\chi'}\spart^2\Psi(\chi') 
+ \frac{1}{\chi}\spart^2 \Psi(\chi) \Psi(\chi') + \frac{1}{\chi} \Psi(\chi) \spart^2 \Psi(\chi') 
\right) \;, \label{new3}
\end{split}
\ee
which come from the coupling between the lens and the time delay that occurs during the longitudinal
photon path.
\item {\em Deflection-deflection} couplings,
\be
4 \frac{\chi_S -\chi}{\chi} \left( 
 2 \spart \dot{\Psi}(\chi) \spart \Psi(\chi') +
 \spart\Psi(\chi) \frac{\chi-\chi'}{\chi'} \spart\Psi(\chi') \right) 
- \frac{8}{\chi}\spart \Psi(\chi) \spart \Psi(\chi')\;, \label{new4}
\ee
which are due to the couplings between two changes in the photon directions. 
Note that the photon deflection is described by a spin-1 field (the deflecting angle), i.e.~a spin raising  operator $\spart$ acting on a scalar. Taken 
alone it does not change the shear as it just affects all photons of the beam in the same way. However,
the coupling of two deflections generate a spin-2 field which contributes to the shear. These corrections, as well as the time delay-lens ones,  are integrated twice along the line of sight.

\end{itemize}
The last line of eq.~(\ref{shear_tot}) contains the effects induced by vector and tensor modes generated at second order. As vectors and tensors are already second-order quantities, they enter linearly in this expression. The vector component enters only through terms integrated along the line of sight. As the tensor component is a spin-2 field it induces also a boundary term, which accounts for the distortion of the shape at the source. The integrated contributions from the tensor component agree with what was found in \cite{Dodelson:2003bv,Sarkar:2008ii}. 
Note that, at second order,  the separation into scalars, vectors and tensors done here is gauge dependent. 
Indeed, we expect all these different contributions to give comparable effects to second-order observables such as the bispectrum. This is similar, for instance, 
to what happens when one computes the CMB bispectrum on large angular scales \cite{Boubekeur:2008kn}.
Finally, note that our final result (\ref{shear_tot}) 
cannot be written as the action of $\spart^2$ on a scalar quantity. 
Thus, the shear will contain also $B$ modes (see Appendix~\ref{app:spart}).

%%%%%%%%%%%%%%%%%%%%%%%%%%%%%%%%%%%%%%%%%%%%%%%%%%%%%%%%%%%%%%%%%%%%%%%%%%%%%%%%%%%%%%%%%%%%%%%%%%%%%%%%%%%%%
%%%%%%%%%%%%%%%%%%%%%%%%%%%%%%%%%%%%%%%%%%%%%%%%%%%%%%%%%%%%%%%%
\subsection{Testing the solution}
\label{sec:test}
%%%%%%%%%%%%%%%%%%%%%%%%%%%%%%%%%%%%%%%%%%%%%%%%%%%%%%%%%%%%%%%%
%%%%%%%%%%%%%%%%%%%%%%%%%%%%%%%%%%%%%%%%%%%%%%%%%%%%%%%%%%%%%%%%%%%%%%%%%%%%%%%%%%%%%%%%%%%%%%%%%%%%%%%%%%%%%

As the reader has certainly realized, the derivation of eq.~(\ref{shear_tot}) is extremely tedious and involves many steps. Thus, it is important to develop tests in order to check this equation and, 
in particular, the new nonlinear couplings (\ref{new1})--(\ref{new4}).
One implicit check is that ${}_2\mD$ behaves as a spin-2 field under rotation of the screen basis. This is automatically ensured by the use of the operators $\spart$ and $\spartb$ instead
of the angular derivatives. 
Furthermore, we can check the terms in eqs.~(\ref{new1})--(\ref{new3}) 
by studying specific cases where part of our calculation can be pursued nonperturbatively. For instance, we can study the shear at linear order in a universe with constant curvature. In the limit where the curvature is small we must recover the couplings between the curvature and the gravitational potential 
given in eqs.~(\ref{new1})--(\ref{new3}). 
As we will see, such a strategy can be generalized to other cases. Unfortunately we were not 
able to develop an analogous test for the terms appearing in eq.~(\ref{new4}).

%%%%%%%%%%%%%%%%%%%%%%%%%%%%%%%%%%%%%%%%%%%%%%%%%%%%%%%%%%%%%%%%%%%%%%%%%%%%%%%%%%%%%%%%%%%%%%%%%%%%%%%%%%%%%
\subsubsection{Couplings from perturbing the spatial curvature}
%%%%%%%%%%%%%%%%%%%%%%%%%%%%%%%%%%%%%%%%%%%%%%%%%%%%%%%%%%%%%%%%%%%%%%%%%%%%%%%%%%%%%%%%%%%%%%%%%%%%%%%%%%%%%

Let us consider a perturbed nonexpanding FLRW spacetime with constant curvature $K$ and metric
\be
\label{fbmetdef}
ds^2=-\big(1+2\phi\big)\dd\eta^2+\frac{1}{\big(1+\psi_K \big)^2} (dr^2+r^2 d\Omega^2)\;,
\ee
with 
\be
\phi =\phi (x^\mu)\,, \qquad \psi_K = \frac{K r^2}{4}\;,
\ee
where we have perturbed only the $00$ part of the metric.
At linear order, the spin-2 mapping field ${}_2 \mD$ for this metric is (see for instance \cite{1997A&A...322....1B})
\be
{}_2 \mD_K(\chi_S) = -   \int_0^{\chi_S} \dd \chi  \frac{D_A(\chi_S-\chi)}{D_A(\chi)} \;   
\spart^2 \phi \;,
\label{D_curved}
\ee
where $D_A (\chi)$ is the angular diameter distance given by eq.~(\ref{dconv}).
Equation (\ref{D_curved}) is just the generalization of eq.~(\ref{2D}), where $D_A(\chi)=\chi$, to a constant curved FLRW universe. 
Since $\psi_K$ does not depend on the angles, $\spart^2 \psi_K =0 $ and thus only $\spart^2 \phi$ appears on the right-hand side of this equation.

For a small curvature $K$ we can expand eq.~(\ref{D_curved}) at first order in $\psi_K$. The angular diameter distance (\ref{dconv}) reads $D_A (\chi) \simeq \chi - K \chi^3 /6$.
Furthermore, we have to evaluate $\phi$ on the geodesic solution for a curved universe, 
i.e.~at $r(\chi) = \chi + K \chi^3 /12$, which yields
\be
\spart^2 \phi (r(\chi),\theta, \phi) = \spart^2 \phi (\chi,\theta, \phi) + \frac{K \chi^3}{12} 
\spart^2 \phi_{, r} (\chi,\theta, \phi) \;.
\ee
Plugging these expressions into eq.~(\ref{D_curved}), replacing the derivative with respect to $r$ using $\partial_r = \dd /\dd \chi - \partial_\chi$ and integrating by parts we obtain, up to first order in $\psi_K$,
\be
{}_2 \mD_K=- \int_0^{\chi_S} \dd \chi  \left[ \frac{\chi_S-\chi}{\chi} \;   
\spart^2 \phi -    \frac{K}{12 } \left(\frac{2\chi_S^3}{\chi}
-3\chi^2-6\chi_S^2+6\chi\chi_S \right)\spart^2\phi -  \frac{K}{12 } (\chi_S-\chi)\chi^2\spart^2\dot{\phi}\right]\;. 
\label{test1}
\ee
The last two terms on the right-hand side of this equation can be seen as ``second-order'' 
corrections to the first-order expression (\ref{2D}), of order $\sim {\cal O} (\phi \psi_K)$, due to the coupling between the gravitational potential $\phi$ and the curvature perturbation $\psi_K$. These corrections are already incorporated 
in our second-order expression (\ref{shear_tot}). Indeed, by replacing $\Psi = (\phi + \psi_K)/2 $ 
in this equation, neglecting second-order terms of order $\sim {\cal O} (\phi^2)$ but keeping those of order 
$\sim {\cal O} (\phi \psi_K)$ one finds, after integrating by parts, eq.~(\ref{test1}).

This calculation can be extended to a spacetime with radial-dependent curvature, i.e. $\psi_K(r)$ a generic function of $r$.
In this case the spin-2 mapping ${}_2 \mD$ is given by
\be
{}_2 \mD_K(\chi_S) =-\int_0^{\chi_S} \dd \chi \frac{G(\chi_S,\chi)}{D_A(\chi)} \spart^2\phi\;,
\label{D_2_green}
\ee
where the Green's function $G(\chi_S,\chi)$ and the angular diameter distance $D_A (\chi)$
can be derived from using eq.~(\ref{D_evol}) in a homogeneous universe with spatial curvature $\psi_K$. 
In this case the trace of eq.~(\ref{D_evol}) becomes 
\be
\frac{\dd^2 D_{A}(\chi)}{\dd\chi^2}=R(r)D_{A}(\chi)\label{fbDAeq}\;,
\ee
where, at first order in $\psi_K$, $R=-{{\psi_K}_{,r}}/{r}-{\psi_K}_{,rr}$, implying
\be
\psi_K=-\int^{\chi}{\dd\chi'}/{\chi'}\int^{\chi'} \dd\chi''\chi''R(\chi'')\; .
\ee

The two solutions of eq.~(\ref{fbDAeq}), $D_{A}(\chi)$ and $C(\chi)$, are determined through their initial conditions, i.e.~$D_{A}(\chi)\to \chi$ and $C(\chi)\to 1+{\cal O}(\chi^2)$, so that 
\bea
D_{A}(\chi)&=&\chi+\int_{0}^{\chi}\dd\chi'\int_{0}^{\chi'}\dd\chi''\,\chi''R(\chi'')\;, \label{D_curved2}\\
C(\chi)&=&1+\int_{0}^{\chi}\dd\chi'\int_{0}^{\chi'}\dd\chi''\,R(\chi'') \;.
\eea
The Green's function is then given by
\be
G(\chi_{S},\chi)=D_{A}(\chi_{S})C(\chi)-D_{A}(\chi)C(\chi_{S})\label{fbGreen} \;.
\ee
Evaluating $\spart^2\phi$ on the geodesic solution for a curved universe, and replacing eqs.~(\ref{D_curved2}) and (\ref{fbGreen}) into (\ref{D_2_green}) we obtain,
at first order in $\phi$ and $\psi_K$ but keeping terms of order $\sim {\cal O} (\phi \psi_K)$,
\be
\begin{split}
{}_2 \mD_K (\chi_S) =&-\ints\spart^2\phi\left[
\frac{\chi_{S}-\chi}{\chi}
\left( 1+
\int_{0}^{\chi}\dd\chi'\int_{0}^{\chi'}\dd\chi''\,R(\chi'')
-\frac{1}{\chi}\int_{0}^{\chi}\dd\chi'\int_{0}^{\chi'}\dd\chi''\,\chi''R(\chi'')
\right)\right.\\
&\left.-
\int_{\chi}^{\chi_{S}}\dd\chi'\int_{0}^{\chi'}\dd\chi''\,R(\chi'')
+\frac{1}{\chi}\int_{\chi}^{\chi_{S}}\dd\chi'\int_{0}^{\chi'}\dd\chi''\,\chi''R(\chi'')
\right]\\
&-\int_{0}^{\chi_{S}}\dd\chi\frac{\chi_{S}-\chi}{\chi}\;\spart^2\phi_{,r}\int_{0}^\chi \psi_{K}\dd\chi'
\;.
\end{split}
\ee
We have checked that, after integration by part of the last line, our general expression (\ref{shear_tot}) reproduces this peculiar case.

%%%%%%%%%%%%%%%%%%%%%%%%%%%%%%%%%%%%%%%%%%%%%%%%%%%%%%%%%%%%%%%%%%%%%%%%%%%%%%%%%%%%%%%%%%%%%%%%%%%%%%%%%%%%%
\subsubsection{Reparametrization invariance under a time shift}
%%%%%%%%%%%%%%%%%%%%%%%%%%%%%%%%%%%%%%%%%%%%%%%%%%%%%%%%%%%%%%%%%%%%%%%%%%%%%%%%%%%%%%%%%%%%%%%%%%%%%%%%%%%%%

Until now we have tested the couplings between the $00$ metric perturbation $\phi(x^\mu)$ and a radial-dependent spatial curvature $\psi_K(r)$. 
Analogously, we can test the nonlinear couplings between the spatial metric perturbation $\psi(x^\mu)$ and a $00$ 
metric perturbation which depends only on time, $\phi_T(\eta)$. Such a perturbation can be reabsorbed into the time coordinate
through a homogeneous shift of the time $\dd \chi = \dd \tilde \chi (1 + \phi_T (\tilde \chi) )$. As a homogeneous time shift does not change the gauge, we expect 
our expression (\ref{shear_tot}) to be invariant under a first-order coordinate change
\be
\chi =  \tilde \chi + \int_0^{\tilde \chi} \dd \tilde \chi' \phi_T\;, \qquad x^i = \tilde x^i \;. \label{time_trans}
\ee

Let us check that this is the case. At first order in $\phi_T$, under this coordinate transformation the metric perturbations change as 
$\phi(\chi) = \tilde \phi(\tilde{\chi})-  \phi_T(\tilde{\chi}) $ and $\psi(\chi) = \tilde \psi(\tilde{\chi}) $. Evaluating $\Psi$ on the geodesic $r =\chi = \tilde \chi + \int_0^{\tilde \chi} \dd \tilde \chi' \phi_T$,
the first term of eq.~(\ref{shear_tot}) transforms as
\be
\begin{split}
-   2 \int_0^{\chi_S} \dd \chi  \frac{\chi_S-\chi}{\chi} \;   
\spart^2 \Psi  =& -   2 \int_0^{\tilde \chi_S} \dd \tilde \chi  \frac{\tilde \chi_S-\tilde \chi}{\tilde \chi} 
\;   \spart^2  \tilde \Psi  \\   
& - 2 \int_0^{\tilde \chi_S} \dd \tilde \chi  \frac1{\tilde \chi} \spart^2 \tilde \Psi \int_0^{\tilde \chi} 
d \tilde \chi' \phi_T +2  \int_0^{\tilde \chi_S} \dd \tilde \chi \frac{\tilde \chi_S-\tilde \chi}{\tilde \chi}
\spart^2\dot{\tilde \Psi}\int_0^{\tilde \chi} d\tilde \chi' \phi_T\;.
\label{detest_t}
\end{split}
\ee
One can check that replacing $\Psi(\chi) = \tilde \Psi(\tilde{\chi})-  \phi_T(\tilde{\chi})$ and $\chi=\tilde{\chi}$ into the 
second-order terms of eq.~(\ref{shear_tot}) exactly cancels the second line of eq.~(\ref{detest_t}),
leaving ${}_2\mD$ invariant under the transformation (\ref{time_trans}).

%%%%%%%%%%%%%%%%%%%%%%%%%%%%%%%%%%%%%%%%%%%%%%%%%%%%%%%%%%%%%%%%%%%%%%%%%%%%%%%%%%%%%%%%%%%%%%%%%%%%%%%%
\subsubsection{Reparametrization under conformal transformation}
\label{sec:conf}
%%%%%%%%%%%%%%%%%%%%%%%%%%%%%%%%%%%%%%%%%%%%%%%%%%%%%%%%%%%%%%%%%%%%%%%%%%%%%%%%%%%%%%%%%%%%%%%%%%%%%%%%

In Appendix~\ref{app:expansion} we show that under the conformal transformation of the metric 
$g_{\mu\nu} \rightarrow \Omega^2 (x^\mu) g_{\mu\nu}$ the mapping $\mD_{ab}$ transforms as 
\be
\mD_{ab} \to \Omega \; \mD_{ab} \;.
\label{conformal_A}
\ee 
If the conformal factor is just $\Omega = 1 + \delta \Omega$, 
where $\delta \Omega$ is a small perturbation 
of order $\Psi$, this conformal transformation is equivalent, at first order in 
$\delta \Omega$, to 
a redefinition of the potentials in the metric,
\be
\psi \rightarrow  \psi-{\delta \Omega}\;, \hspace{0.4cm} \phi\rightarrow \phi+{\delta \Omega}~.
\label{phi_psi_weyl}
\ee
Since, as expected, the Weyl potential 
$\Psi=(\phi+\psi)/2$ does not change under this transformation, in eq.~(\ref{shear_tot}) only the boundary term
proportional to $\psi(\chi_S)$ is not invariant. Transforming this term according to (\ref{phi_psi_weyl}) yields a contribution to  ${}_2 \mD(\chi_S)$ in the new metric,
\be
{}_2 \mD(\chi_S) \to {}_2 \mD(\chi_S)  - 2 \delta \Omega(x^\mu_S) \int_0^{\chi_S} d\chi \frac{\chi_S-\chi}{\chi}\spart^2\Psi\;.
\ee
This is exactly what we expect from the 
conformal transformation ${}_2 \mD \to (1+ \delta \Omega) {}_2 \mD$.

%%%%%%%%%%%%%%%%%%%%%%%%%%%%%%%%%%%%%%%%%%%%%%%%%%%%%%%%%%%%%%%%%%%%%%%%%%%%%%%%%%%%%%%%%%%%%%%%%%%%%%%%
%%%%%%%%%%%%%%%%%%%%%%%%%%%%%%%%%%%%%%%%%%%%%%%%%%%%%%%%%%%%%%%%%%%%%%%%%%%%%%%%%%%%%%%%%%%%%%%%%%%%%%%%
\subsection{The reduced shear}
\label{sec:reduced}
%%%%%%%%%%%%%%%%%%%%%%%%%%%%%%%%%%%%%%%%%%%%%%%%%%%%%%%%%%%%%%%%%%%%%%%%%%%%%%%%%%%%%%%%%%%%%%%%%%%%%%%%
%%%%%%%%%%%%%%%%%%%%%%%%%%%%%%%%%%%%%%%%%%%%%%%%%%%%%%%%%%%%%%%%%%%%%%%%%%%%%%%%%%%%%%%%%%%%%%%%%%%%%%%%

As mentioned in Sec.~\ref{sec:weak},
the quantity that we measure is the reduced shear, which is given by the ratio between 
the spin-2 anisotropic mapping ${}_2\mD$ and the trace $\mD$, eq.~(\ref{reduced}).
Expanding this equation up to second order using that $\mD = 2 \chi_S + \delta \mD$  we obtain for the reduced shear
\be
 g  = - \frac{{}_2 \mD}{2 \chi_S} + \frac{{}_2 \mD \; \delta \mD}{(2 \chi_S)^2}  \;,
\label{g}
\ee
where ${}_2\mD$ in the first term on the right-hand side is given by eq.~(\ref{shear_tot}) and, from eqs.~(\ref{2D}) and (\ref{traceD}), the second-order correction on the right-hand side is given by
\be
\begin{split}
  \frac{{}_2 \mD \; \delta \mD}{(2 \chi_S)^2}  =&   \int_0^{\chi_S} d\chi \frac{\chi_S-\chi}{\chi \chi_S}\spartb\spart \Psi \int_0^{\chi_S} d\chi' 
\frac{\chi_S-\chi'}{\chi' \chi_S}\spart^2 \Psi  \\
& +\psi(\chi_S) \int_0^{\chi_S}d\chi \frac{\chi_S-\chi}{\chi \chi_S}\spart^2 \Psi
-2 \int_0^{\chi_S} d\chi \Psi  \int_0^{\chi_S} d\chi' \frac{\chi_S-\chi'}{\chi' \chi_S^2}\spart^2 \Psi \;. 
\label{g_2}
\end{split}
\ee
The first line is the usual correction to the reduced shear 
due to the coupling between the convergence and the shear \cite{Dodelson:2005rf,Schneider:1997ge}. 
The two corrections in the second line are negligible in the small-angle approximation. 
The term proportional to $\psi(\chi_S)$ comes from the coupling between the lens and the curvature perturbation
at the source contained in the trace of $\mD_{ab}$, see eq.~(\ref{traceD}).
Note that it cancels with the 
one in the expression of ${}_2 \mD $. Indeed, as both the isotropic and anisotropic part of 
$\mD_{ab}$ change according to eq.~(\ref{conformal_A}) 
under conformal transformation, we expect the reduced shear, which is their ratio, 
to depend only on the Weyl potential $\Psi$, which is invariant under conformal transformation.
The last term in the second line is the coupling between the time delay contained in $\mD$ with the lens.
These terms are thus of the same order as those discussed in eqs.~(\ref{new1})--(\ref{new3})

The expression of the reduced shear above is given taking the source 
at constant conformal time $\eta_S$. 
However, in order to relate this quantity to observations we need to compute it
using a constant redshift $z_S$ for the source, given by
\be
z_S=
\frac{{k}_S^\al {v}_{S\al}}{ {k}_O^\al {v}_{O\al}} - 1\;.
\label{redshift_total}
\ee 
As in the Poisson gauge the $z=$ const hypersurfaces do not
coincide with the $\eta=$ const hypersurfaces, the redshift $z_S$ is not homogeneous and
we expect a correction to eq.~(\ref{g}) coming from the coupling between
the perturbed redshift plane of the source and the lens.
As explained in Sec.~\ref{sec:weak}, 
eq.~(\ref{redshift_correction}), the reduced shear at constant redshift is given by
\be
g_z = g + \delta g_z \;, \qquad \delta g_z \equiv  -\frac{\dd g}{\dd \chi_S} \; \frac{\dd \chi_S }{\dd z_S} \; \delta z_S\;,
\label{g_correction}
\ee
where $\delta z_S $ is the perturbation of the redshift (\ref{redshift_total}).
We must now reintroduce the expansion of the Universe.
By taking the derivative with respect to $\chi_S$ of the linear expression for $g$, eq.~(\ref{shear_first}), we find
\be
\frac{\dd g}{\dd \chi_S}  \; \frac{\dd \chi_S }{\dd z_S} = \frac1{\chi_S^2 H_S} \ints \spart^2 \Psi\;,
\label{expr1}
\ee
where $H$ is the Hubble rate defined from the cosmic time $t$, $\dd t = - a \dd \chi$, 
as $H\equiv \frac{1}{a}\frac{\dd a}{\dd t }$.

To compute $\delta z_S$ we can perturb at first order eq.~(\ref{redshift_total}) 
using the expression for $k^0$ given in eq.~(\ref{k^0}). Setting to zero 
the perturbations of the metric and of the velocity at the observer position, we obtain
\be
\delta z_S=- (1+ z_S)\left(\phi(\chi_S)+\ve_r|_{\chi_S}\cdot{\bf v}_S  - 2 \int_0^{\chi_S}\dd\chi  \dot \Psi  \right)\;.
\label{redshift_pert}
\ee
In the three terms on the right-hand side of this equation 
one recognizes the Sachs-Wolfe, the Doppler and the integrated Sachs-Wolfe effects contributing 
to the photon redshift perturbation. 
Finally, combining the expression (\ref{expr1}) and the redshift perturbation (\ref{redshift_pert}) we obtain for the redshift correction (\ref{g_correction}),
\be
\delta g_z =  \frac{1 +z_S}{\chi_S^2 H_S} 
 \left(\phi(\chi_S) +\ve_r|_{\chi_S}\cdot{\bf v}_S - 2 \int_0^{\chi_S}\dd\chi  \dot \Psi  \right) \ints \spart^2 \Psi \;.
\label{redshift_correction2}
\ee

The observed reduced shear at constant redshift becomes then\footnote{Note that the only effect of the transverse velocity of the source is to modify the direction under which the galaxy is observed. Thus, it has no effect on the shear.}   
\be
g_z = - \frac{{}_2 \mD}{2 \chi_S} + \frac{{}_2 \mD \; \delta \mD}{(2 \chi_S)^2}  + \delta g_z \;,
\label{final_result}
\ee
where the first, the second and the third terms on the right-hand side are respectively given by eqs.~(\ref{shear_tot}), (\ref{g_2}) and~(\ref{redshift_correction2}). 
This is the main result of this article.

\section{Conclusion}
\label{sec:conclusion}
In this article we have derived the expression of the reduced cosmic shear up to second order in the perturbations with full-sky validity. 
Our main result is summarized in eq.~(\ref{final_result}). As it is expressed in terms of spin-2 operators on the sphere it can be decomposed as sum of spin-weighted spherical harmonics on the sky. Indeed, this description ensures that our observable has a genuine spin-2 behavior on the celestial sphere.

Our result is written in terms of the metric perturbations in the generalized Poisson gauge. These are the scalar potentials $\phi$ and $\psi$ and the vector and tensor components of the metric generated at second order, respectively $\omega_i$ and $h_{ij}$. Let us first comment on the first two terms on the right-hand side of eq.~(\ref{final_result}).
Remarkably, the contribution from scalar perturbations from the sum of these two terms
can be expressed in terms of the Weyl potential $\Psi = (\phi+\psi)/2$ only. As explained, this is due to the fact that null geodesics are conformally invariant. 
These two terms contain the well-known second-order corrections due to lens-lens coupling and departure from the Born approximation, which dominate in the small-angle approximation. On larger angular scales new couplings become important. These are an intrinsic contribution which is a purely general relativistic effect at second order, a coupling between the gravitational potential at the source with the lens and corrections due to couplings between the lens and the photon time-delay. We have checked that these contributions can be independently reconstructed from the calculation of the shear at first order in a universe with a radially dependent spatial curvature. Other checks, such as the invariance under a homogeneous time shift and a conformal transformation can be used to verify the validity of these new corrections. Another scalar correction appears in the form of products of two spin-1 fields and comes from the couplings between two photon deflections. Finally, besides the scalar contributions, the shear gets a contribution from  spin-2 quantities defined from the vector and tensor components of the metric generated at second-order. Note that the separation between all these contributions is not gauge invariant.

In Poisson gauge, the correction due to the coupling between the photon redshift perturbation and the lens cannot be written in terms of $\Psi$ only. Indeed, the integrated contribution to the photon redshift -- the integrated Sachs-Wolfe effect -- is a time integral over $\dot \Psi$ but the intrinsic contributions -- Sachs-Wolfe and Doppler effects -- are expressed in terms of the Newtonian gravitational potential and the velocity along the line of sight and do not depend on $\Psi$ only.

We are now in the position to explore the phenomenological consequences of these results in view of the future (partially) full-sky lensing surveys. In particular, the new corrections that we have computed
should become relevant in deriving the lensing bispectrum on large angular scales. For instance, to compute the bispectrum in the squeezed limit one needs to take one of the three modes to be much smaller than the other two, corresponding to angular scales comparable to the depth of the survey. 
As the lensing is a cumulative effect integrated along the line of sight, it is difficult, at this stage, to precisely guess the relative importance of the various contributions. In particular, although the lens-lens coupling terms are a priori larger by a factor $\sim l^2$ compared to the others, they may be damped by geometrical factors. We leave these investigations for the future.

\section*{Ackowledgements}
It is a pleasure to thank Chiara Caprini for useful discussions and collaboration in the early stage of the preparation of this article and Nicolas Van de Rijt for spotting few typos. C.B.~and F.V.~acknowledge support from the EU Marie Curie Research
and Training network ÓUniverseNetÓ (MRTN-CT-2006-035863).

\vspace{1cm}

\hrule 

\vspace{0.3cm}

%\begin{widetext}
\appendix

\section{Conformal relations}
 \label{app:expansion}
 
We consider two conformally related metrics $\tilde{g}_{\al\beta}=\Omega^2(x^\mu) g_{\al\beta}$.  We want to show that the corresponding Jacobi mappings computed using 
eq.~(\ref{Devol}) with the respective metrics 
are related by $\tilde{\mD}_{ab} =\Omega \; \mD_{ab}$. 
We can decompose the Riemann tensor in terms of the Ricci tensor $R_{\alpha \beta}$ 
and the Weyl tensor $C_{\al\beta\rho\sigma}$ as
\cite{wald}
\be
R_{\al\beta\gamma \delta} = C_{\al\beta\gamma \delta}+g_{\al[\gamma} R_{\delta ]\beta} - g_{\beta [\gamma} R_{\delta ] \alpha} - \frac16 R g_{\alpha [ \gamma} g_{\delta ] \beta}\;,
\ee
where the brackets in the indices denote antisymmetrization. 
With this decomposition, the definition of $\mR_{ab}$, eq.~(\ref{R_def}), simply yields
\be
\label{conformal_transf_a}
\mR_{ab} = C_{\al\beta\gamma \delta} n_a^{\ \alpha} k^\beta k^\gamma n_b^{\ \delta} - \frac12 \delta_{ab} R_{\beta \gamma} k^\beta k^\gamma \;,
\ee
where we have used the normalization properties of $n_a^{\ \mu}$ and $k^\mu$. 

Using this expression, let us study how $\mR_{ab}$ transforms under a conformal
transformation. As
$n_a^{\ \mu}$ is normalized to unity it transforms as $\tilde n_a^{\ \mu}= 
\Omega^{-1} 
n_a^{\ \mu}$. If $\lambda$ is an affine parameter of the null geodesic 
in the metric $g_{\alpha \beta}$, then the affine parameter computed using the metric 
$\tilde g_{\alpha \beta} $ is related to $\lambda$ by $\dd \tilde \lambda = \Omega^2 
\dd \lambda$ \cite{wald}. Thus, $\tilde k^{\mu}= \Omega^{-2} k^{\mu}$. Furthermore,
as the Weyl tensor with one upper index $C_{\alpha \beta \delta}{}^{\gamma}$
is invariant under conformal transformation, the first term of 
(\ref{conformal_transf_a}) transforms as
\be
\tilde C_{\al\beta\gamma \delta} \tilde n_a^{\ \alpha} \tilde k^\beta 
\tilde k^\gamma \tilde n_b^{\ \delta} = \Omega^{-4} C_{\al\beta\gamma \delta} n_a^{\ \alpha} k^\beta k^\gamma n_b^{\ \delta} \;.
\ee
The Ricci tensor transforms as \cite{wald}
\be
\tilde R_{\alpha \gamma} = R_{\alpha \gamma} -2 \nabla_\alpha \nabla_\gamma \ln \Omega
- g_{\alpha \gamma} g^{\delta \sigma} \nabla_{\delta} \nabla_\sigma \ln \Omega + 2
(\nabla_\alpha \ln \Omega) \nabla_{\gamma} \ln \Omega - 2 
g_{\alpha \gamma} g^{\delta \sigma} (\nabla_\delta \ln \Omega) \nabla_\sigma \ln \Omega\;.
\ee
Projecting this expression by $\tilde k^\alpha \tilde k^\gamma$, the two terms proportional to $g_{\alpha \gamma}$ vanish because of the null condition of the photon wave vector, 
while the covariant derivatives can be written as derivatives along the null geodesic.
Using the geodesic equation one obtains
\be
\tilde R_{\beta \gamma} \tilde k^\beta \tilde k^\gamma = \Omega^{-4} \left[ 
R_{\beta \gamma} k^\beta k^\gamma 
-2  \frac{\dd^2 \ln \Omega}{\dd \lambda^2} 
+  2 \left( \frac{\dd \ln \Omega}{\dd \lambda}\right)^2 \right]\;.
\ee
Thus, from eq.~(\ref{conformal_transf_a}) $\mR_{ab}$ transforms as
\be
\tilde{\mR}_{ab}= \Omega^{-4} \left[  \mR_{ab}
+  \frac{\dd^2 \ln \Omega}{\dd \lambda^2} \delta_{ab}
- \left( \frac{\dd \ln \Omega}{\dd \lambda}\right)^2 \delta_{ab} \right]\;.
\ee
Finally, 
using this transformation it is easy to show that if $\mD_{ab}$ is the solution of
\be
\frac{\dd^2}{\dd\lambda^2}\mD_{ab} =\mR_{ac} \mD_{cb}\;,
\ee
then $\tilde{\mD}_{ab} = \Omega \; \mD_{ab}$ is the solution of the corresponding equation for the metric $\tilde{g}_{\mu\nu}$,
\be
\frac{\dd^2}{\dd\tilde{\lambda}^2}\tilde{\mD}_{ab}=\tilde{\mR}_{ac} \tilde{\mD}_{cb}\;.
\ee
The relation between $\mD_{ab}$ and $\tilde{\mD}_{ab}$ can be easily understood by noting that $\mD_{ab}$ relates distances at the source, that scale like $\Omega$, to angles at the observer that are invariant under a conformal transformation. 

\section{Spin operators on the sphere}

\label{app:spart}

The construction of spin fields can be easily done on a plane identified with the complex plane of coordinates $z=x+\ii y$.
Let us consider a complex field ${}_s X(z)$ whose value depends on $z$. This field will be said to have spin-\textit{s} if its value is changed in $e^{\ii s \alpha} {}_s X(z)$ after a rotation of angle $\alpha $. For instance, if ${}_0 X(z)$ is a scalar (i.e. a spin-0) field, then
\begin{equation}
{}_1 X(z)={\partial_x} \; {}_0 X(z)+\ii{\partial_y}\; {}_0 X(z)\; ,
\end{equation}
is a spin-1 field. This relation can alternatively be written as
${}_1 X(z)=2{\partial_z}\;  {}_0 X(z)$,
where the partial derivative is to be taken for a fixed value of $\zb$. In general, the successive application of the operator 
$\spart\equiv 2{\partial_z}$
leads to the construction of spin-\textit{s} fields. Equivalently, the operator
$\spartb\equiv 2{\partial_{\zb}}$
lowers the spin by one. In the context of standard first-order lensing theory, the complex shear field $\gamma=\gamma_{1}+\ii \gamma_{2}$ is a spin-2 field that derives from the projected potential $\psi$, i.e. $\gamma=\spart^2\psi$. 

This construction can be extended to the sphere when one does not want to rely on the small-angle approximation. The early elements of such a construction date back to~\cite{NewmanPenrose66,Goldberg1967}. In general, one is naturally led to introduce the Euler angles  $(\theta,\varphi,\alpha)$ so that the coordinates of a point on the unit sphere are 
$(\sin\theta\,\cos\varphi,\sin\theta\,\sin\varphi,\cos\theta)$. To each of these points one can associate a radial vector $\ve_r$ and two tangential vectors $\ve_{1}$ and $\ve_{2}$ that can be conveniently chosen along the lines $\varphi={\rm const}$ and $\theta={\rm const}$ respectively if $\alpha=0$. The angle $\alpha$ then corresponds to a rotation around the axis $\ve_r$ that rotates $\ve_{\theta}$ and $\ve_{\varphi}$ with an angle $\alpha$.  

As for the plane, a spin-\textit{s} field is such that its phase varies as $s\alpha$, by rotation of an angle $\alpha$. The construction of the operators $\spart$ and $\spartb$ relies on the use of the complex stereographic coordinates,
\begin{equation}
\ze=\cot\left(\frac{\theta}{2}\right)\,\exp(\ii\varphi)\; ,
\end{equation}
which map the sphere onto the complex plane. More specifically it can be shown that the expressions of the operators $\spart$ and $\spartb$ depend explicitly on the spin of the field ${}_s X$ to which they are applied, 
\begin{equation}
\spart\,  {}_sX=2P^{1-s}\ {\partial_{\ze}}P^s {}_s X\quad \hbox{and}\quad 
\spartb\, {}_sX=2P^{1+s}\ {\partial_{\zeb}}P^{-s}{}_s X\; ,
\end{equation}
where 
$P=\frac{1}{2}(1+\ze\zeb)$.
Expressing these operators in terms of $\theta$ and $\varphi$, one finds
\begin{equation}
\label{spart}
\spart \, {}_sX=-(\sin \theta)^s\left(\partial_\theta+\ii\csc \theta \partial_\varphi \right)(\sin \theta)^{-s}{}_s X\ \ \hbox{and}\ \ 
\spartb\, {}_sX=-(\sin \theta)^{-s}\left(\partial_\theta-\ii\csc \theta\partial_\varphi \right)(\sin \theta)^{s}{}_s X\; .
\end{equation}
In analogy to the case of the plane, the Laplacian operator formally reads
$\Delta=\spartb\spart$. Note however that the relation $\spartb\spart=\spart\spartb$ holds only when the operators act on a scalar field. In general we have, $(\spartb\spart-\spart\spartb){}_s X=2s {}_s X$.

Spherical harmonics, $\ylm$, are spin-$0$ functions that are the eigenfunctions of the Laplacian with eigenvalue $-l(l+1)$, i.e.
$\Delta\ylm=-l(l+1)\ylm$
with
${\partial_\varphi}\ylm=\ii m\,\ylm$
and with a specific normalization. The orthogonality relation,
\begin{equation}
\int\dd^2\Omega\ \ylmp\ \ylmc=\delta_{l'l}\,\delta_{m'm}\label{orthnYlm}\; ,
\end{equation}
is the key property that makes it possible to decompose any function into spherical harmonics.

In general spin-\textit{s} fields are decomposed on the basis of the spin-weighted spherical harmonics, that can be obtained through the application of the operator $\spart$ and $\spartb$ on the spherical harmonics. More specifically, we define $\yslm$ with
\begin{eqnarray}
\yslm&=&\left[\frac{(l-s)!}{(l+s)!}\right]^{1/2}\spart^s\ylm,\quad (0\le s\le l)\; ,\\
\yslm&=&\left[\frac{(l+s)!}{(l-s)!}\right]^{1/2}(-1)^s\spartb^{-s}\ylm,\quad (-l\le s\le 0)\; .
\end{eqnarray}
The spin-weighted spherical harmonics obey the following relations,
\begin{eqnarray}
\yslmc&=&(-1)^{m}\ymslmm\; , \\
\spartb\spart \yslm&=&-(l-s)(l+s+1)\yslm \; , \\
\int\dd^2\Omega\ \yslmp\ \yslmc&=&\delta_{l'l}\,\delta_{m'm}\label{orthnsYlm} \; , \\
\sum_{lm}\yslmvp\yslmc&=&\delta(\varphi-\varphi')\delta(\cos\theta-\cos\theta')\; ,
\end{eqnarray}
with
$\int\dd^2\Omega\equiv\int_{0}^{2\pi}d\varphi\int_{-1}^{1}d\cos \theta$.

One usually defines two scalars, $E$ and $B$, associated to the spin-2 shear $\gamma$. To do that, one decomposes the shear $\gamma$ and its complex conjugate $\gamma^*$ as,
 \begin{equation}
\gamma(\theta,\varphi)=\sum_{lm}\,_{2}a_{lm}\,_{2}\ylm\ \quad \hbox{and}\ \quad \gamma^{*}(\theta,\varphi)=\sum_{lm}\,_{-2}a_{lm}\,_{-2}\ylm \; .
\end{equation}
As in the context of CMB polarization (see \cite{1997PhRvD..55.1830Z}), $E$ and $B$ can be defined through their harmonic decomposition,
\begin{equation}
E(\theta,\varphi)=-\frac{1}{2}\sum_{lm}\left(\,_{2}a_{lm}+\,_{-2}a_{lm}\right)\,\ylm\ \quad \hbox{and}\ \quad B(\theta,\varphi)=\frac{\ii}{2}\sum_{lm} \left(\,_{2}a_{lm}-\,_{-2}a_{lm}\right)\,\ylm.
\end{equation}
$E$ is invariant under parity change, whereas $B$ changes signs.

\bibliography{Lens2}

\end{document}